\documentclass[aoas,MSNbibl,nameyear,dvips]{arximspdf}
\usepackage{graphicx}
\doi{10.1214/13-AOAS685} 
\volume{8}
\issue{1}
\pubyear{2014}
\firstpage{19}
\lastpage{47}

\makeatletter
\newcommand{\rrVert}{\Vert}
\newcommand{\llVert}{\Vert}
\newcommand{\eqref}[1]{(\ref{#1})}

\makeatother

\begin{document}
\begin{frontmatter}

\title{Hierarchical array priors for ANOVA decompositions of
cross-classified data\thanksref{T1}}
\thankstext{T1}{Supported in part by NICHD Grant 1R01HD067509-01A1.}
\runtitle{Array priors for ANOVA}
\pdftitle{Hierarchical array priors for ANOVA decompositions of
cross-classified data}

\begin{aug}
\author[a]{\fnms{Alexander} \snm{Volfovsky}\ead
[label=e1]{volfovsky@fas.harvard.edu}}
\and
\author[b]{\fnms{Peter D.} \snm{Hoff}\corref{}\ead[label=e2]{pdhoff@uw.edu}}
\affiliation{Harvard University and University of Washington}
\runauthor{A. Volfovsky and P. Hoff}
\address[a]{Department of Statistics\\
Harvard University\\
1 Oxford St\\
Cambridge, Massachusetts 02138\\
USA\\
\printead{e1}}

\address[b]{Department of Statistics\\
University of Washington\\
Box 354322\\
Seattle, Washington 98195\\
USA\\
\printead{e2}}
\end{aug}

\received{\smonth{8} \syear{2012}}
\revised{\smonth{9} \syear{2013}}

%
\begin{abstract}
ANOVA decompositions are a standard method for describing and estimating
heterogeneity
among the means of a response variable across
levels of multiple categorical
factors. In such a decomposition,
the complete set of main effects and interaction terms can
be viewed as a collection of vectors, matrices and arrays
that share various index sets defined by the
factor levels.
For many types of categorical factors,
it is plausible that an ANOVA decomposition
exhibits some consistency across orders of effects, in that
the levels of a factor that
have similar main-effect coefficients
may also have similar coefficients in higher-order interaction terms.
In such a case, estimation of the higher-order interactions
should be improved by borrowing information from the main effects and
lower-order interactions.
To take advantage of such patterns,
this article
introduces
a class of hierarchical prior distributions for collections of interaction
arrays that can adapt to the presence of such interactions.
These prior distributions are
based on a type of array-variate normal distribution, for which
a covariance matrix for each factor is estimated.
This prior is able to adapt to potential similarities
among the levels of a factor, and incorporate any such information into
the estimation of the effects in which the factor appears.
In the presence of such similarities,
this prior is able to borrow information from well-estimated main
effects and lower-order interactions to assist in the estimation of
higher-order terms for which data information is limited.
\end{abstract}

%
\begin{keyword}
\kwd{Array-valued data}
\kwd{Bayesian estimation}
\kwd{cross-classified data}
\kwd{factorial design}
\kwd{MANOVA}
\kwd{penalized regression}
\kwd{tensor}
\kwd{Tucker product}
\kwd{sparse data}
\end{keyword}

\pdfkeywords{Array-valued data,
Bayesian estimation,
cross-classified data,
factorial design,
MANOVA,
penalized regression,
tensor,
Tucker product,
sparse data}

\end{frontmatter}

\section{Introduction}\label{sec1}
Cross-classified data are prevalent
in many disciplines,
including the
social and health sciences.
For example, a survey or observational study
may record health behaviors of its participants,
along with a variety of
demographic variables, such as
age, ethnicity and education
level, by which the participants can be classified.
A common data analysis goal in such settings is the estimation
of the health behavior means for each combination of levels
of the demographic factors. In a three-way layout, for example,
the goal is to estimate the three-way table of population cell
means, where each cell corresponds to a particular combination
of factor levels. A standard estimator of the table is provided
by the table of sample means, which can alternatively be
represented by its ANOVA decomposition into additive
effects and two- and three-way interaction terms.

The cell sample means
provide an unbiased estimator of the population means,
as long as there are observations available for each cell.
However, if the cell-specific sample sizes are small, then it may be
desirable to share information across the cells to reduce the variance
of the estimator. Perhaps the simplest and most common
method of information sharing
is to assume that certain mean contrasts among levels of one set of factors
are equivalent
across levels of another set of factors or, equivalently, that
certain interaction terms in the ANOVA decomposition
of population cell means are exactly zero.
This is a fairly large modeling assumption,
and can often be rejected
via plots or standard $F$-tests.
If such assumptions are rejected,
it still may be desirable to share information across
cell means, although perhaps in
a way that does not
posit exact relationships among them.

%
\begin{table}
\tabcolsep=0pt
\caption{Cross-tabulation of the sample sizes for the demographic
variables. ``Hispanic'' is coded as ``Hispanic, not Mexican.'' The
categories of education are as follows: P-Primary, S-Secondary, HD-High
school degree, AD-Associate's degree, BD-Bachelor's degree}
\label{Flo:dat_sum}
\begin{tabular*}{\textwidth}{@{\extracolsep{\fill}}lcccccccccccccccccccc@{}}
\hline
&\multicolumn{5}{c}{\textbf{Mexican}}&\multicolumn{5}{c}
{\textbf{Hispanic}}
&\multicolumn{5}{c}{\textbf{White}}&\multicolumn{5}{c@{}}{\textbf
{Black}}\\[-6pt]
&\multicolumn{5}{c}{\hrulefill}&\multicolumn{5}{c}
{\hrulefill}
&\multicolumn{5}{c}{\hrulefill}&\multicolumn{5}{c@{}}{\hrulefill}\\
\textbf{Age} & \textbf{P} & \textbf{S} & \textbf{HD} & \textbf{AD}
& \textbf{BD} & \textbf{P} & \textbf{S} & \textbf{HD} & \textbf
{AD} & \textbf{BD} &
\textbf{P} & \textbf{S} & \textbf{HD} & \textbf{AD} &
\textbf{BD} & \textbf{P} & \textbf{S} & \textbf{HD} & \textbf{AD}
& \textbf{BD} \\
\hline
31--40 & 21 & 24 & 23 & 17 & 13 & 12 & 8 & 10 & 11 & \phantom{0}1 &
\phantom{0}3 & 37 & 56 & 55 & 56 & \phantom{0}1 & 13 & 31 & 35 & 16 \\
41--50 & 26 & 10 & 19 & 14 & \phantom{0}6 & 11 & 9 & 10 & \phantom
{0}9 & \phantom{0}3 & 10 & 25 & 56 & 57 & 50 & \phantom{0}2 & 25 & 21
& 25 & 17 \\
51--60 & 29 & 11 & 10 & 14 & 10 & 17 & 6 & 12 & 13 & 11 & 10 & 24 & 46
& 57 & 57 & \phantom{0} & 23 & 23 & 24 & 14 \\
61--70 & 31 & \phantom{0}7 & \phantom{0}5 & 11 & \phantom{0}5 & 19 &
4 & 11 & \phantom{0}6 & \phantom{0}7 & 15 & 23 & 56 & 46 & 54 & 16 &
34 & 20 & 33 & 14 \\
71--80 & 27 & \phantom{0}2 & \phantom{0}3 & \phantom{0}1 & \phantom
{0}3 & 10 & 8 & \phantom{0}5 & \phantom{0}2 & \phantom{0}7 & 61 & 37
& 93 & 72
& 68 & 16 & 10 & 11 & \phantom{0}7 & 12 \\
\hline
\end{tabular*}
\end{table}


As a concrete example, consider
estimating mean
macronutrient intake across
levels of age (binned in 10 year increments), ethnicity and education from
the National Health and Nutrition Examination Survey
(NHANES).
Table~\ref{Flo:dat_sum}
summarizes the cell-specific sample sizes
for intake of overall carbohydrates as well as two subcategories (sugar
and fiber)
by age, ethnicity and education levels for male respondents
(more details on these data are provided in Section~\ref{sec:Data-analysis}).
%
Studies of carbohydrate intake have been motivated by a frequently
cited relationship
between carbohydrate intake and health outcomes [\citet
{chandalia2000beneficial,moerman1993dietary}].
Studies of obesity in the US have shown an overall increase
in caloric intake primarily due to an increase in carbohydrate
intake from 44 to 48.7 percent of total calories from 1971 to 2006
[\citet{austin2011trends}].
Recently, the types of carbohydrates 
that are being consumed have become
of primary interest.
For example, in the study of cardiovascular disease,
simple sugars are associated with raising triglycerides
and overall cholesterol while dietary fiber has been
associated with lowering triglycerides [\citet
{albrink1986interaction,yang2003carbohydrate}].
Total carbohydrates and the types of carbohydrates have also been targeted
in recent studies of effective weight loss [e.g.,
sugar consumption in the form of HFCS in drinks, \citet{nielsen2004changes}].

However, these studies generally report on marginal means of carbohydrate
intake across demographic variables,
and do not take into account potential nonadditivity,
or interaction terms, between them
[\citet
{park2011dietary,montonen2003whole,basiotis1989sources,verly2010sources,johansson2001underreporting}].
In a study where nonadditivity was considered, the authors
only tested for the presence of a small subset of possible interactions
and did not consider any interactions of more than two effects [\citet
{austin2011trends}].
A more detailed understanding of the relationship between mean
carbohydrate intake and
the demographic variables can be obtained from a MANOVA decomposition
of the means array into
main-effects, two- and three-way interactions.
Evidence for interactions for multivariate data can be assessed with
approximate $F$-tests based on the Pillai trace statistics
[\citet{olson1976choosing}].

%
\begin{table}
\caption{MANOVA testing of interaction terms via
Pillai's trace statistic}
\label{Flo:manova-table}
\begin{tabular*}{\textwidth}{@{\extracolsep{\fill}}lcccc@{}}
\hline
& \multicolumn{1}{c}{\textbf{approx} $\bolds{F}$} & \multicolumn
{1}{c}{\textbf{num df}} &
\multicolumn{1}{c}{\textbf{den df}} & \multicolumn{1}{c@{}}{$\bolds
{p}$\textbf{-value}} \\
\hline
Education & 11.15 & \phantom{0}15 & 6102 & $<$0.01 \\
Ethnicity & 18.07 & \phantom{00}9 & 6102 & $<$0.01 \\
Age & 21.38 & \phantom{0}12 & 6102 & $<$0.01 \\
Education:Ethnicity & \phantom{0}1.67 & \phantom{0}36 & 6102 &
\phantom{0.}0.01 \\
Education:Age & \phantom{0}1.60 & \phantom{0}48 & 6102 & \phantom
{0.}0.01 \\
Ethnicity:Age & \phantom{0}2.05 & \phantom{0}36 & 6102 & $<$0.01 \\
Education:Ethnicity:Age & \phantom{0}1.44 & 144 & 6102 & $<$0.01 \\
\hline
\end{tabular*}
\end{table}


%
\begin{figure}[t]
\includegraphics{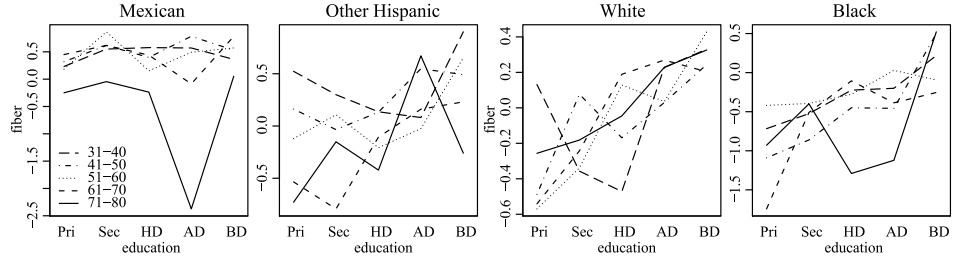}
\caption{Three-way interaction plot of fiber cell means by ethnicity,
age and education level.}
\label{Flo:three-way-OLS}
\end{figure}

For our data, the
$F$-tests presented in
Table~\ref{Flo:manova-table}
indicate strong evidence
that the two- and three-way interactions are not zero.
Based on these results, standard practice would be
to retain the full model and
describe the interaction patterns via
various contrasts of cell sample means.
Often this is done by visual examination of
interaction plots, that is, plots of cell means by
various combinations of factors.
For example, Figure~\ref{Flo:three-way-OLS}
gives the
age by education interaction plots for
each of the four ethnicity groups. The three-way interaction
between ethnicity, age and education can be described as the
inconsistency of the
two-way interactions
across levels of ethnicity. Visually, there is some indication
that Mexican respondents have a different age by education interaction
than the other ethnicities,
but it is difficult
to say anything more specific.
Indeed, it is difficult to even describe the two-way interactions,
due to the high variability of the cell sample means.

Much of the heterogeneity in these plots can
be attributed to the low sample sizes in many cells and the
resulting
sampling variability of the cell sample means.
A cleaner picture of the three-way interactions could
possibly be obtained via cell mean estimates with lower variability.
A variety of
penalized least squares procedures have been proposed in order to
reduce estimate variability and mean squared error (MSE),
such as ridge regression and the lasso.
Recent variants of these approaches allow for
different penalties on ANOVA terms of different orders,
including the ASP method of \citet{beran2005asp}
and grouped versions of the lasso
[\citet{yuan2007model,friedman2010note}].
Corresponding Bayesian approaches include
Bayesian
lasso procedures [\citet{yuan2005efficient,genkin2007large,park2008bayesian}]
and multilevel hierarchical priors
[\citet
{pittau2010economic,park2006state,hodges2007smoothing,cui2010partitioning}].

While these procedures attain a reduced MSE by shrinking
linear model coefficient estimates toward zero,
they do not generally take full advantage of the structure
that is often present in cross-classified data sets.
In the data analysis example above, two of the three factors (age and
education) are ordinal,
with age being a binned version of a continuous predictor.
Considering factors such as these more generally,
suppose a categorical factor $x$ is a binned version of some underlying
continuous or ordinal explanatory variable $\tilde x$ (such as
income, age, number of children or education level).
If the mean of the response variable $y$
is smoothly varying in the underlying variable $\tilde x$,
we would expect that adjacent levels of the factor $x$ would have
similar main effects and interaction terms.
Similarly, for nonordinal factors (such as ethnic group or religion)
it is possible that two levels represent similar populations, and thus
may have similar main effects and interaction terms as well.
We refer to such
similarities across the orders of the effects as
\emph{order consistent interactions}.



%
\begin{figure}

\includegraphics{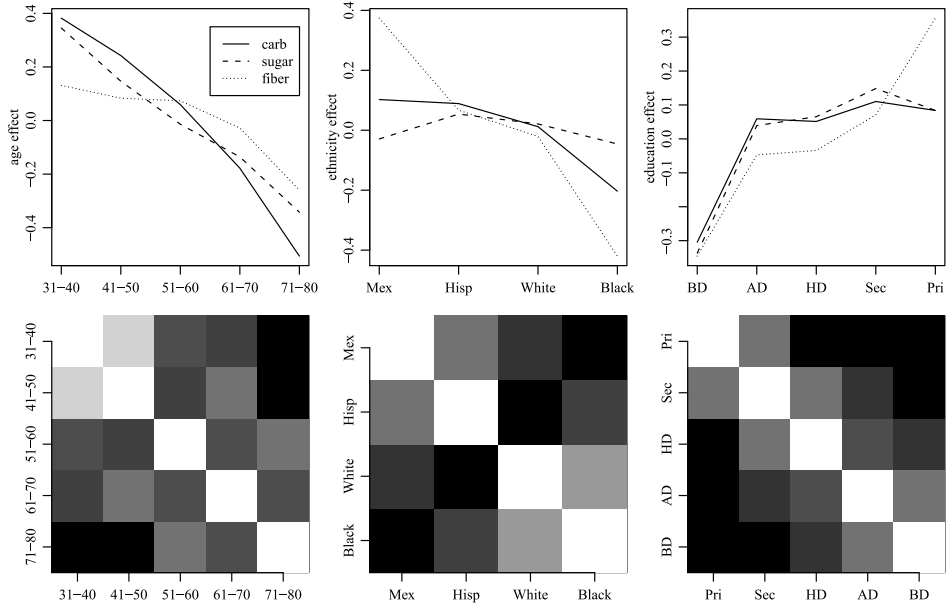}

\caption{Plots of main effects and interaction correlations for the three
outcome variables (carbohydrates, sugar and fiber). The first row of plots
gives OLS estimates of the
main effects for each factor. The second row of plots
gives correlations of effects between levels of each factor, with
white representing 1 and black representing $-1$. The interactions are
calculated
based on OLS estimates of the main effects and two-way interactions
of each factor.}
\label{Flo:two-way-manova}
\end{figure}

Returning to the NHANES data,
Figure~\ref{Flo:two-way-manova} summarizes the
OLS
estimates of the
main effects and
two-way
interactions
for the
three outcome variables (carbohydrates, sugar and fiber).
Not surprisingly, the main effects
for the ordinal factors (age and education) are ``smooth,'' in
that the estimated main effect for a given level
is generally similar to the effect for an adjacent level.
Additionally, some similarities among the ethnic groups
appear consistent across the three outcome variables.
To assess consistency of such similarities
between main effects and two-way interactions,
we computed correlations of parameter estimates for these
effects between levels
of each factor. For example,
there are $3\times10 =30$ main-effect and two-way interaction estimates
involving each level of age: For each of the three outcome variables,
there is 1 main-effect estimate for each age level,
4 estimates from the age by ethnicity interaction and 5 estimates
from the age by education interaction.
We compute a correlation
matrix for the five levels of age
based on the resulting $30\times5$ matrix of parameter estimates, and
similarly compute
correlations among levels of ethnicity and among levels of
education.
The second row of Figure~\ref{Flo:two-way-manova}
gives grayscale plots of these correlation matrices. The results
suggest some degree of order consistent interactions:
For the ordinal factors, the highest correlations
are among adjacent pairs. For the ethnicity factor,
the results suggest that, on average, the effects for the Mexican
category are more similar to the Hispanic
(not Mexican) 
category than to the
other ethnic categories, as we might expect.



The OLS estimates of the main effects
and three-way interactions
presented above, along with the
fact that two of the three factors are ordinal,
suggest the possibility of
order consistent interactions among the array of population cell means.
More generally,
order consistent interactions may be present
in a variety of data sets encountered in the social and health sciences,
especially those that include ordinal factors, or factors
for which some of the levels may represent very similar populations.
In this paper, we propose a novel class of hierarchical prior distributions
over main effects and interaction arrays that
can adapt to the presence of order consistent interactions.
The hierarchical prior distribution provides joint estimates of a
covariance matrix
for each factor, along with the factor main effects and interactions.
Roughly speaking,
the covariance matrix for a given factor is estimated from the main
effects and interactions in which the factor appears.
Conversely, an estimate of a factor's covariance matrix can assist in
the estimation of higher-order interactions, for which data information is
limited.
We make this idea more formal in the next section, where we construct
our
prior distribution from a set of related array normal distributions
with separable covariance structures [\citet{hoff2011separable}] and
provide a Markov chain Monte Carlo algorithm for inference under this
prior.
In Section~\ref{sec:Simulation-study} we provide a simulation study
comparing estimation under
our proposed prior to some standard estimators. As expected, our
approach outperforms others when the data exhibit order consistent
interactions. Additionally, for data lacking any interactions, our approach
performs comparably to the OLS estimates obtained from the additive
model (i.e., the oracle estimator).
In Section~\ref{sec:Data-analysis} we extend this methodology to
MANOVA models in order to
analyze the multivariate NHANES data presented above.
In addition to estimates of main effects and interactions, our
analysis provides measures of similarity between levels of each of the factors.
We conclude in Section~\ref{sec5} with a summary of our approach and
a discussion of possible extensions.


\section{A hierarchical prior for interaction arrays}\label{sec:Model}

In this section we introduce the hierarchical array (HA) prior
and present a Markov chain Monte Carlo (MCMC) algorithm for posterior
approximation and parameter estimation.
The HA prior is constructed from several
semi-conjugate priors, and so the MCMC algorithm can be based on
a straightforward Gibbs sampling scheme.


\subsection{The hierarchical array prior}\label{sec2.1}
For notational convenience we consider the case of three
categorical factors, and note that the HA prior generalizes trivially
to accommodate a greater number
of factors.
Suppose the three categorical factors
have levels $ \{ 1,\ldots,m_{1} \} $,
$ \{ 1,\ldots,m_{2} \} $ and $ \{ 1,\ldots,m_{3} \}$,
respectively.
The standard ANOVA model for a
three-way factorial data set
is
\begin{eqnarray}
\label{eq:modelbalcomp} y_{ijkl} 
&=& \mu+a_{i}+b_{j}+c_{k}+
(ab )_{ij}+ (ac )_{ik}+ (bc )_{jk}+ (abc
)_{ijk} + \varepsilon_{ijkl},
\nonumber
\\[-8pt]
\\[-8pt]
\nonumber
\{ \varepsilon_{ijkl} \} &\sim& \mbox{i.i.d. normal} \bigl(0,
\sigma^{2} \bigr).
\nonumber
\end{eqnarray}
Let $a$ denote the $m_1\times1$ vector
of main effects for the first factor,
$(ab)$ denote the $m_1\times m_2$ matrix describing the two-way interaction
between the first two factors, $(abc)$ denote the
$m_1\times m_2\times m_3$ three-way interaction array,
and let $b$, $c$, $(ac)$ and $(bc)$ be defined similarly.
Bayesian inference for this model proceeds by specifying a
prior distribution for the
ANOVA decomposition $\theta= \{ \mu,a,b,c,(ab), (ac), (bc),(abc) \}$ and
the error variance $\sigma^2$.

As described in the \hyperref[sec1]{Introduction}, if two levels of a
factor represent
similar populations, we would expect that coefficients of the decomposition
involving these two levels would have similar values. For example,
suppose levels $i_1$ and $i_2$ of the first factor correspond to similar
populations. We might then expect $a_{i_1}$ to be close to $a_{i_2}$,
the vector
$\{ (ab)_{i_1,j}, j=1,\ldots, m_2\}$ to be close to the vector
$\{ (ab)_{i_2,j}, j=1,\ldots, m_2\}$, and so on. We represent this potential
similarity between levels of the first factor with
a covariance matrix $\Sigma_a$, and consider a mean zero prior
distribution on
the ANOVA decomposition such that
%
\begin{eqnarray*}
\operatorname{Cov} [ a ] &=& \mathrm{E} \bigl[aa^T \bigr] =
\Sigma_a,
\\
\mathrm{E} \bigl[ (ab) (ab)^T \bigr] &=& k_{ab}
\Sigma_a,
\\
\mathrm{E} \bigl[ (ac) (ac)^T \bigr] &=& k_{ac}
\Sigma_a,
\\
\mathrm{E} \bigl[ (abc)_{(1)} (abc)_{(1)}^T \bigr]
&=& k_{abc} \Sigma_a,
\end{eqnarray*}
where $k_{ab}$, $k_{ac}$ and $k_{abc}$ are scalars.
Here, $(abc)_{(1)}$ is the \emph{matricization} of the array
$(abc)$, which converts the $m_1\times m_2\times m_3$ array into
an $m_1\times(m_2 m_3)$ matrix by adjoining the $m_3$ matrices
of
dimension $m_1\times m_2$ that form the array $(abc)$
[\citet{kolda2009tensor}].
To accommodate similar structure for the second and third factors,
we propose the following prior covariance model for the
main effects and interaction terms:
\begin{eqnarray*}
\operatorname{Cov }[ a ] &=& \Sigma_a,\qquad\operatorname{Cov }[ b
] = \Sigma_b,\qquad\operatorname{Cov }[ c ] = \Sigma_c
\\
\operatorname{Cov } \bigl[ \operatorname{vec}(ab) \bigr]& =& \Sigma
_b \otimes \Sigma_a /\gamma_{ab},\qquad
\operatorname{Cov } \bigl[ \operatorname{vec}(bc) \bigr] = \Sigma_c
\otimes\Sigma_b/\gamma_{bc},
\\
\operatorname{Cov } \bigl[ \operatorname{vec}(ac) \bigr] &=& \Sigma
_c \otimes \Sigma_a /\gamma_{ac} \qquad
\operatorname{Cov } \bigl[ \operatorname{vec}(abc) \bigr] = \Sigma _c
\otimes \Sigma_b \otimes\Sigma_a/\gamma_{abc},
\end{eqnarray*}
where ``$\otimes$'' is the Kronecker product.
The covariance matrices $\Sigma_a$, $\Sigma_b$ and $\Sigma_c$
represent the similarities between the levels of each of the three
factors, while the scalars $\gamma_{ab},\gamma_{ac},\gamma
_{bc},\gamma_{abc}$
represent the relative (inverse) magnitudes of the interaction terms as
compared to the main effects.
Further specifying the priors on the ANOVA decomposition parameters
as being mean-zero and Gaussian, the prior on $a$
is then the multivariate normal distribution
$N_{m_1}( 0,\Sigma_a)$,
and the prior on $\operatorname{vec}(ab)$ is
$N_{m_1m_2}( 0, \Sigma_b \otimes\Sigma_a/\gamma_{ab})$.
This latter distribution is sometimes referred to as a
matrix normal distribution [\citet{dawid1981some}].
Similarly, the prior on $\operatorname{vec} (abc)$ is
$N_{m_1m_2m_3}( 0,\Sigma_c \otimes\Sigma_b \otimes\Sigma_a/\gamma_{abc})$,
which has been referred to as an array normal distribution
[\citet{hoff2011separable}].

In classical ANOVA decompositions, it is common to impose an identifiability
constraint on the different effects. In a Bayesian analysis it is
possible to place priors over identifiable sets of parameters,
but this is cumbersome and not frequently done in practice
[\citet{gelman2007data,kruschke2010doing}]. The priors
we propose for the effects in the ANOVA decomposition
in this article
induce a prior over the cell means, which are identifiable.
These priors have an intuitive interpretation and do not
negatively affect the convergence of MCMC
chains generated by the proposed procedure
as can be seen in the
Simulation and Application sections.

In most data analysis situations
the similarities between the levels of a given factor
and magnitudes of the interactions relative to the main effects
will not be known in advance.
We therefore consider a hierarchical
prior so that $\Sigma_a$, $\Sigma_b$, $\Sigma_c$
and the $\gamma$-parameters
are estimated from the data.
Specifically, we use independent
inverse-Wishart prior distributions for each covariance matrix,
for example, $\Sigma_a \sim$ inverse-Wishart$( \eta_{a0}, S_{a0}^{-1})$,
and gamma priors for the $\gamma$-parameters, for example, $\gamma
_{ab} \sim\operatorname{gamma}(\nu_{ab0}/2,\tau^2_{ab0}/2)$, where
$\eta_a$, $S_a$, $\nu_{ab0}$ and $\tau^2_{ab0}$ are hyperparameters
to be specified (some default choices for these parameters are
discussed at the end of this section).
This hierarchical prior distribution can be viewed as an adaptive
penalty, which allows for sharing of information across
main effects and interaction terms.
For example, estimates of the
three-way interaction
will be stabilized by the
covariance matrix
$\Sigma_c\otimes\Sigma_b\otimes\Sigma_a$,
which in turn is influenced by
similarities between levels of
the factors that are consistent across the main effects, two-way and
three-way interactions.


\subsection{Posterior approximation}\label{sub:fullcond}
Due to the semi-conjugacy of the HA prior,
posterior approximation can be obtained from
a straightforward
Gibbs sampling scheme.
Under this scheme, iterative simulation of
parameter values from the corresponding full conditional distributions
generates a Markov chain having a stationary distribution equal
to the target posterior distribution.
For computational simplicity, we consider the case of a balanced data set
in which the sample size in each cell is equal to some common value $n$,
in which case
the data can be expressed as an $m_1 \times m_2 \times
m_3 \times n$ four-way array $Y$.
A modification of the algorithm to accommodate unbalanced data is
discussed in the next subsection.

Derivation of the full conditional distributions of the
grand mean $\mu$ and the
error variance $\sigma^2$
are completely standard:
Under a $N(\mu_0,\tau_0^2)$ prior for~$\mu$,
the corresponding full conditional distribution is
$N(\mu_1,\tau^2_1)$, where
$\tau_1^2 = ( 1/\tau_0^2 + n m_1 m_2 m_3 /\sigma^2 )^{-1} $ and
$\mu_1 = \tau_1^2 (\mu_0/\tau_0^2 + n m_1 m_2 m_3 \bar r/\sigma^2)$,
where $\bar r = \sum_{ijkl} ( y_{ijkl}- [ a_i +b_j + c_k + (ab)_{ij} +
(ac)_{ik} + (bc)_{jk} + (abc)_{ikj} ] )/n$.
Under an inverse-$\operatorname{gamma} (\nu_{0}/2,\nu_{0}\sigma
_{0}^{2}/2 )$
prior distribution,
the full conditional distribution of $\sigma^2$ is
an inverse-$\operatorname{gamma} ({\nu}_{1}/2,{\nu}_{1}{\sigma
}_{1}^{2}/2 )$
distribution, where
${\nu}_{1}=\nu_{0}+n m_1 m_2 m_3$, ${\nu}_{1}{\sigma}_{1}^{2}=\nu
_{0}\sigma_{0}^{2}+\sum_{ijkl} (y_{ijkl}-\mu_{ijk} )^{2}$
and $\mu_{ijk}= \mu+ a_i +b_j + c_k + (ab)_{ij} +
(ac)_{ik} + (bc)_{jk} + (abc)_{ikj}$.
Derivation of the full conditional distributions of parameters other
than $\mu$ and $\sigma^2$
is straightforward, but slightly nonstandard due to the
use of matrix and array normal prior distributions for the
interaction terms.
In what follows, we compute the full conditional
distributions for a few of these parameters. Full conditional distributions
for the remaining parameters can be derived in an analogous fashion.


\textit{Full conditionals of $a$ and $(abc)$.}

To identify the full conditional distribution of the vector $a$
of main effects for the first factor, let
\begin{eqnarray*}
r_{ijkl} & = & y_{ijkl}- \bigl(\mu+b_{j}+c_{k}+
(ab )_{ij}+ (ac )_{ik}+ (bc )_{jk}+ (abc
)_{ijk} \bigr)
\\
& = & a_{i}+\varepsilon_{ijkl},
\end{eqnarray*}
that is, $r_{ijkl}$ is the ``residual'' obtained by subtracting all
effects other than $a$ from
the data.
Since $\{ \varepsilon_{ijkl}\} \sim \mathrm{i.i.d.}\ \operatorname{normal}(0,\sigma^2)$,
we have
\[
p \bigl( Y | \theta, \sigma^2 \bigr) \propto_a \exp
\biggl\{ -\frac{m_2 m_3 n}{2\sigma^2} \bigl( a^T a - 2a^T \bar r
\bigr) \biggr\},
\]
where
$\bar r = (\bar r_1,\ldots, \bar r_{m_1}) $ with $\bar r_i=\sum_{jkl}
r_{ijkl}/(m_2 m_3 n) $,
$\theta= \{ \mu, a,b,c,(ab),(ac),\break(bc),(abc) \}$ and
``$\propto_a$'' means ``proportional to as a function of $a$.''
Combining this with the $N_{m_1}(0,\Sigma_a)$ prior density
for $a$, we have
\[
p \bigl( a | Y, \theta_{-a}, \sigma^2,
\Sigma_{a} \bigr) \propto_a \exp \biggl(-\frac{m_{2}m_{3}n}{2\sigma^{2}}
\bigl[a^{T}a-2a^{T} \bar{r} \bigr]-\frac{1}{2}a^{T}
\Sigma_{a}^{-1}a \biggr)
\]
and so the full conditional distribution of $a$ is multivariate normal
with
\begin{eqnarray*}
\operatorname{Var} \bigl[ a | Y,\theta_{-a},\sigma^2,
\Sigma_{a} \bigr] &=& \bigl(\Sigma_{a}^{-1}+I
m_{2}m_{3}n/ \sigma^{2} \bigr)^{-1},
\\
\operatorname{Exp} \bigl[a | Y,\theta_{-a},\sigma^2,
\Sigma_{a} \bigr] &=& \bigl(\Sigma_{a}^{-1}+I
m_{2}m_{3}n/ \sigma^{2} \bigr)^{-1} \bar
r \times \bigl(m_2m_3 n /\sigma^2 \bigr),
\end{eqnarray*}
where $I$ is the $m_1\times m_1$ identity matrix.

Derivation of the full conditional distributions for the interaction
terms is similar.
For example, to obtain the full conditional distribution of $(abc)$,
let $r_{ijkl}$ be the residual obtained after subtracting all
other components of $\theta$ from the data, so that
$r_{ijkl}= (abc )_{ijk}+\varepsilon_{ijkl}$.
Let $\bar r$ be the three-way array of cell means of $\{ r_{ijkl} \}$,
so that
$\bar{r}_{ijk}=\sum_{l}r_{ijkl}/n$.
Combining the likelihood in terms of $\bar r$ with
the $N_{m_1m_2m_3}( 0, \Sigma_{c}\otimes\Sigma_{b}\otimes\Sigma
_{a}/\gamma_{abc})$ prior density for $\operatorname{vec}(abc)$ gives
\begin{eqnarray*}
&&p \bigl( (abc )|Y,\sigma^{2},\Sigma_{a},
\Sigma_{b},\Sigma_{c},\gamma_{abc},
\theta_{-(abc)} \bigr)
\\
&&\qquad\propto_{(abc)} \exp \biggl(- \frac{n}{2} \bigl[
\operatorname{vec} (abc )^{T}\operatorname{vec} (abc )-2
\operatorname{vec} (abc )^{T}\operatorname{vec}(\bar{r}) \bigr] \biggr)
\\
& &\qquad\quad{}\times\exp \biggl(-\frac{1}{2}\operatorname{vec} (abc
)^{T} (\Sigma _{c}\otimes\Sigma_{b}\otimes
\Sigma_{c}/ \gamma_{abc} )^{-1}\operatorname{vec}
(abc ) \biggr)
\end{eqnarray*}
and so $\operatorname{vec}(abc)$ has a multivariate normal
distribution with
variance and mean given by
\begin{eqnarray*}
&&\operatorname{Var}\bigl[ \operatorname{vec}(abc) | Y,\theta _{-(abc)},
\sigma^2, \Sigma_{a},\Sigma_{b},
\Sigma_{c}, \gamma_{abc} \bigr] \\
&&\qquad= \bigl( (
\Sigma_{c}\otimes \Sigma_{b}\otimes\Sigma_{a}/
\gamma_{abc} )^{-1}+I n/\sigma^{2}
\bigr)^{-1},
\\
&&\mathrm{E}\bigl[ \operatorname{vec}(abc) | Y,\theta_{-(abc) },
\sigma^2, \Sigma_{a},\Sigma_{b},
\Sigma_{c}, \gamma_{abc} \bigr] \\
&&\qquad=\bigl( (
\Sigma_{c}\otimes \Sigma_{b}\otimes\Sigma_{a}/
\gamma_{abc} )^{-1} +I n/\sigma^{2}
\bigr)^{-1} \operatorname{vec}(\bar{r}) \times n/\sigma^{2}.
\end{eqnarray*}
Full conditional distributions for the remaining effects can be derived
analogously.


\textit{Full conditional of $\Sigma_a$.}
The parameters in the ANOVA decomposition whose priors depend on
$\Sigma_a$ are $a$, $(ab)$, $(ac)$ and $(abc)$.
For example, the prior
density of $(ab)$ given $\Sigma_a$, $\Sigma_b$ and $\gamma_{ab}$
can be written as
\begin{eqnarray*}
p \bigl((ab)| \Sigma_a,\Sigma_b,\gamma_{ab}
\bigr) & = & | 2\pi\Sigma_b \otimes\Sigma_a/
\gamma_{ab} |^{-1/2} \\
&&{}\times\exp \bigl( -\operatorname{vec}(ab)^T
[ \Sigma_b \otimes\Sigma_a/\gamma_{ab}
]^{-1} \operatorname{vec}(ab)/2 \bigr)
\\
&\propto_{\Sigma_a}& | \Sigma_a|^{-m_2/2}
\operatorname{etr} \bigl( - \Sigma_a^{-1}
\gamma_{ab} (ab)^T \Sigma_b^{-1}
(ab)/2 \bigr)
\\
& =& |\Sigma_a |^{-m_2/2} \operatorname{etr} \bigl(-
\Sigma_{a}^{-1} S_{ab}/2 \bigr),
\end{eqnarray*}
where $S_{ab} = \gamma_{ab}(ab)^T \Sigma_b^{-1} (ab)$ and
$\operatorname{etr}(A) = \exp\{ \operatorname{trace}(A) \}$ for a
square matrix $A$.
Similarly, the priors for $a$, $(ac)$ and $(abc)$
are proportional to
$|\Sigma_a|^{-d_i/2}\times\break \operatorname{etr} (-\Sigma_a^{-1} S_i/2 )$
(as a function of $\Sigma_a$) for $i\in\{a,ac,abc \}$
where
%
\begin{eqnarray*}
S_a &=& a a^T,
\\
S_{ac} &= & \gamma_{ac} (ac)^T
\Sigma_c^{-1} (ac),
\\
S_{abc} &=& \gamma_{abc} (abc)_{(1)} (
\Sigma_c \otimes\Sigma_b )^{-1}
(abc)_{(1)},
\end{eqnarray*}
and $d_a = 1$, $d_{ac}=m_3$ and $d_{abc}=m_2 m_3$.
The inverse-Wishart$(\eta_{a0},S_{a0}^{-1} )$ prior density
for $\Sigma_a$ can also be written in a similar fashion: it is proportional
to $|\Sigma_a|^{-(\eta_{a0}+m_1+1)/2}\operatorname{etr} (-\Sigma_a^{-1}
S_{a0}/2 )$.
Multiplying together the prior densities for
$a$, $(ab)$, $(ac)$, $(abc)$
and
$\Sigma_a$ and simplifying by the additivity of exponents and the
linearity of
the trace gives
\begin{eqnarray*}
p( \Sigma_{a} | \theta,\Sigma_b,\Sigma_c,
\gamma)& \propto&\vert\Sigma_{a}\vert^{- (1+m_1 + \eta_{a0}
+1+m_{2}+m_{3}+m_{2}m_{3} )/2}\\
&&{}\times \operatorname{etr}
\bigl( -\Sigma_a^{-1} (S_{a0} + S_a +
S_{ab} + S_{ac} + S_{abc} )/2 \bigr).
\end{eqnarray*}
It follows that
the full conditional distribution of $\Sigma_a$ is
inverse-Wishart$ (\eta_{a1},S_{a1}^{-1} )$, where
$\eta_{a1} = \eta_{a0}+ (1+m_{2}+m_{3}+m_{2}m_{3} )$ and
$S_{a1} = S_{a0} + S_{a}+ S_{ab}+ S_{ac} + S_{abc}$.
The full conditional expectation of $\Sigma_{a}$ is therefore
$S_{a1}/ (\eta_{a1}-m_{1}-1 )$,
which combines
several estimates of the
similarities among the levels of the first factor, based
on the main effects and the interactions.

\textit{Full conditional of $\gamma_{abc}$.}
The full conditional distribution of $\gamma_{abc}$ depends only
on the $ (abc )$ interaction term.
The normal prior for $(abc)$ can be written as
\begin{eqnarray*}
&&p \bigl( (abc) | \Sigma_a,\Sigma_b,
\Sigma_c,\gamma_{abc} \bigr)\\
&&\qquad\propto_{\gamma
_{abc}}
\gamma_{abc}^{m_1m_2 m_3/2} \exp \bigl\{ -\gamma_{abc}
\operatorname {vec}(abc)^T [ \Sigma_c\otimes
\Sigma_b \otimes\Sigma_a ]^{-1}
\operatorname{vec}(abc)^T/2 \bigr\}.
\end{eqnarray*}
Combining this density with
a $\operatorname{gamma}(\nu_{abc0}/2, \tau^2_{abc0}/2 )$ prior density
yields a full
conditional for $\gamma_{abc}$ that is $\operatorname{gamma} (\nu
_{abc1}/2,\tau
^2_{abc1}/2 )$,
where
\begin{eqnarray*}
\nu_{abc1} & = & \nu_{abc0}+m_{1}m_{2}m_{3},\label{eq:gampostn}
\\
\tau^2_{abc1} & = & \tau^2_{abc0}+
\operatorname{vec}(abc)^T [ \Sigma_c\otimes
\Sigma_b \otimes\Sigma_a ]^{-1}
\operatorname{vec}(abc).
\end{eqnarray*}

\subsection{Balancing unbalanced designs}\label{sec2.3}

For most survey data we expect the sample sizes $\{ n_{ijk}\}$
to vary across combinations of factors.
As a result, the full conditional distributions
of the ANOVA decomposition parameters are more
difficult to compute. For example,
the conditional variance of the three-way interaction $\operatorname{vec}(abc)$
changes from $ (\gamma_{abc} (\Sigma_{c}\otimes\Sigma_{b}\otimes
\Sigma_{a} )^{-1}+I n/\sigma^{2} )^{-1}$ in the balanced case to
$ (\gamma_{abc} (\Sigma_{c}\otimes\Sigma_{b}\otimes\Sigma
_{a} )^{-1}+D /\sigma^{2} )^{-1}$
in the general case,
where $D$ is a diagonal matrix with diagonal
elements $\operatorname{vec}(\{ n_{ijk} \}$).
Even for moderate numbers of levels of the factors, the
matrix inversions required to calculate
the full conditional distributions in the unbalanced case can slow down
the Markov
chain considerably.
As an alternative, we propose the following data augmentation procedure to
``balance'' an unbalanced design.
Let $\bar Y^o$ be the
three-way array of cell means based on the observed data,
that is, $\bar y_{ijk}^o = \sum y_{ijkl}/n_{ijk}$.
Letting $n=\max(\{ n_{ijk}\} )$,
for each cell $ijk$ with sample size $n_{ijk}<n$ and
at each step of the Gibbs sampler, we impute a
cell mean based on the ``missing'' $n-n_{ijk}$ observations as
$\bar{y}_{ijk}^m \sim\operatorname{normal}(\mu_{ijk},\sigma
^{2}/[n_{\max}-n_{ijk}])$,
where $\mu_{ijk}$ is the population mean for cell $ijk$ based on
the current values of the ANOVA decomposition parameters.\vspace*{-1pt}
We then combine $\bar y_{ijk}^o$ and $\bar y_{ijk}^m$ to form
the ``full sample'' cell mean $\bar y_{ijk}^f =
(n_{ijk} \bar y_{ijk}^o + (n-n_{ijk}) \bar y_{ijk}^m )/n $.
This array of cell means provides the sufficient statistics
for a balanced data set, for which the full conditional distributions
derived above can be used.

\subsection{Setting hyperparameters}\label{sec2.4}
In the absence of detailed prior information about the parameters,
we suggest using
a modified empirical Bayes approach to hyperparameter selection based
on the maximum likelihood estimates (MLEs) of the
error variance and
mean parameters. Priors for $\mu$ and $\sigma^2$ can be set as
unit information priors [\citet{kasswasserman1995}], whereby
hyperparameters are chosen so that the
prior means are near the MLEs but the prior variances
are set to correspond roughly to only one observation's worth of information.
For the covariance matrices
$\Sigma_a$, $\Sigma_b$ and $\Sigma_c$, recall that
the prior for the main effect $a$ of the first factor is
$N_{m_1}( 0,\Sigma_a)$. Based on this, we choose the prior
for $\Sigma_a$ to be
inverse-Wishart$(\nu_{a0}, S_{a0}^{-1})$ with
$\nu_{a0}=m_{1}+2$ and $S_{a0}=\|\hat a\|^2I_{m_{1}}/m_1$,
where $\hat a$ is the MLE of $a$ and $\| \hat a\|$ is the $L_2$ norm
of $\hat a$.
Under this prior, $\mathrm{E}[ \operatorname{tr}(\Sigma_a) ] = \|
\hat a\|^2$,
and so the scale of the prior matches the empirical estimates.
Finally, the $\gamma$-parameters can be set analogously,
using diffuse gamma priors but centered around values
to match the magnitude of the OLS estimates of the interaction terms
they correspond
to, relative to the magnitude of the main effects.
For example, in the next section we use
a
$\operatorname{gamma}(\nu_{ab0}/2, \tau^2_{ab0}/2)$ prior for
$\gamma_{ab}$ in which
$\nu_{ab0}=1$ and
$\tau_{ab0}^2= \|\hat a\|^2 \|\hat b\|^2 /\|\hat{ (ab) } \|^2$,
where $\hat a$, $\hat b$ and $\hat{ (ab) } $ are the OLS estimates.

The above procedure can be modified to accommodate an
incomplete design, where
not all the OLS estimates are available for a complete model.
For example, in a two-way example, if exactly one cell is empty,
then the OLS estimates are available for all effect levels
except for the two-way interaction
for the missing cell. Abusing notation a bit, let $\|(\hat{ab})\|$ be
the $L_2$ norm
of available OLS estimates for the two-way interaction. There are $m_1
m_2 -1$
of these.
Note that this will likely underestimate
$\|(ab)\|$, as it is missing the component contributed by the
missing cell. To correct for this underestimate, we propose
the following modification for setting the hyperparameters:
$\|(\tilde{ab})\|^2 = \|(\hat{ab})\|(m_1 m_2)/(m_1 m_2 -1)$.
The choice of $\tau^2_{ab0}$ above becomes $\|\hat a\|^2\|\hat b\|^2/\|
(\tilde{ab})\|^2$.



\section{Simulation study}\label{sec:Simulation-study}

This section presents the results of 
four simulation studies comparing
the HA prior to several
competing approaches. The first simulation study uses data generated
from a means array that exhibits order consistent
interactions.
Estimates based on the HA prior outperform standard OLS estimates
as well as estimates from
a standard Bayesian (SB) approach as in \citet{gelman2005analysis},
and is also related to a grouped version of the lasso procedure
[\citet{yuan2006model}].
The second simulation study
uses data from a means array that exhibits ``order inconsistent''
interactions, that is, interactions without consistent similarities in
parameter values
between levels of a factor.
In this case the HA prior still outperforms
the OLS and standard Bayes approaches, although not by as much as
in the presence of order consistent interactions.
In the third simulation we study the Bayes risk of the HA
procedure when data is generated directly from the SB prior.
Unlike the second simulation study, where interactions were ``order
inconsistent''
but had potential similarities, in this case all effects were completely
independent and so the oracle SB approach that imposes independence
on the interaction effects outperforms HA, though not by much.
The
fourth simulation study
uses data from a means array that has an exact additive decomposition,
that is, there are no interactions. The HA prior procedure
again outperforms the standard Bayes and OLS approaches, although it does
not do as well as
OLS and Bayes oracle estimators that assume the
correct additive model.

The Markov chain Monte Carlo algorithms were implemented using the
{\sf R} statistical programming language on a computer with a 2.5 GHz
processor.
The additive Bayes approach is significantly faster
than the other two Bayesian procedures since it contains the fewest
parameters. The other two procedures are comparable, but with SB
being somewhat faster than HA on average.
Specifically,
for the simulations conducted below, SB ran an estimated 17\% faster
than HA, which had a run time on the order of 16 minutes per data set
(depending on
sample size). The overall runtime improves by almost 50\% if the data set
is balanced.

%
\begin{figure}

\includegraphics{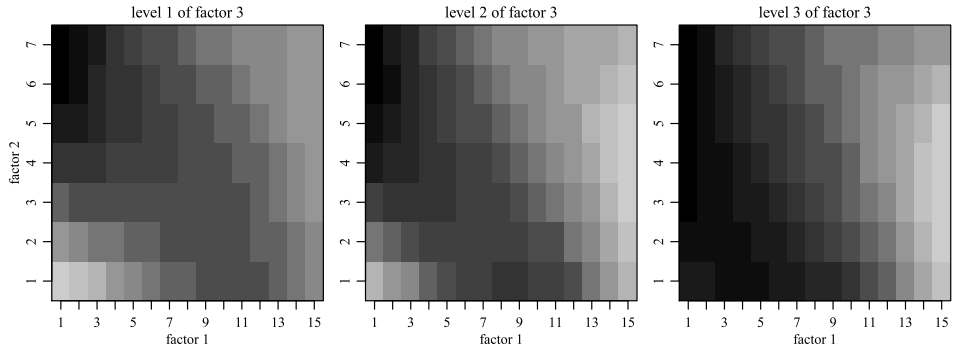}

\caption{The means array $M$ across levels of the third factor.}
\label{Flo:marray_full}
\end{figure}

\subsection{Data with order consistent interactions}\label
{sub:Data-with-interactions}

The data in this simulation study is generated from a model where the means
array exhibits order consistent interactions. The dimensions
of the means array $M$ were chosen to be $m_{1}\times m_{2}\times
m_{3}=15\times7\times3$,
which could represent, for example, the number of categories we might
have for age, education level and political affiliation in a cross-classified
survey data set. The means array was generated
from a cubic function of three variables that was then binned. %
Figure~\ref{Flo:marray_full} plots the mean array
across the third factor, demonstrating the nonadditivity present
in $M$. By decomposing $M$ into the main, two-way and three-way
effects in the same manner as described in Section~\ref{sec:Model},
we can summarize the nonadditivity of $M$ through the magnitudes
of the different sums of squares. The magnitudes of the main effects,
given by the squared $L_2$ norm of the effects,
$\llVert a\rrVert^{2}/m_1,\llVert b\rrVert^{2}/m_2\mbox
{ and}\llVert
c\rrVert^{2}/m_3$,
are $5.267,0.012,0.004$, respectively. Those of the two-way interactions
$\llVert ab\rrVert^{2}/(m_1 m_2),\llVert ac\rrVert
^{2}/(m_1 m_3)\mbox{ and}\llVert bc\rrVert^{2}/(m_2 m_3)$
are $1.365,1.312\mbox{ and }0.384$, and the magnitude
of the three-way interaction $\llVert abc\rrVert^{2}/(m_1 m_2
m_3)$ is
$0.474$.
For each sample size $\{400, 1000, 5000, 10\mbox{,}000\}$,
we simulated 50 data sets using the mean array $M$ and
independent standard normal errors.
In order to make a comparison to
OLS possible, we first allocated one observation to each cell of the
means array. We then distributed the remaining observations uniformly
at random (with replacement) among the cells of the means array.
This leads to a complete but potentially unbalanced design.
The average number of observations per cell under the
sample sizes
$\{400,1000,5000,10\mbox{,}000\}$
was $\{1.3, 3.2, 15.9, 31.7\}$.

For each simulated data set we obtained estimates under the HA prior
(using the hyperparameter specifications described in Section~\ref{sec2.4}),
as well as
ordinary least squares estimates (OLS) and posterior estimates under
a standard Bayesian prior (SB).
The SB approach is essentially a simplified version of the HA prior
in which the parameter values are conditionally independent given the
hyperparameters:
$ \{ a_{i} \} \sim\mathrm{i.i.d.}\ \mathrm{N} (0,\sigma_{a}^{2} )$,
$ \{ (ab )_{ij} \} \sim\mathrm{i.i.d.}\ \mathrm{N} (0,\sigma
_{ab}^{2} )$
and $\{ (abc )_{ijk}\} \sim\mathrm{i.i.d.} (0,\sigma
_{abc}^{2} )$,
and similarly for all other main effects and interactions.
To facilitate comparison to the HA prior,
the hyperpriors
for these $\sigma^2$-parameters
are the same as
the hyperpriors for the inverses of the $\gamma$-parameters in the HA
approach. As a result, this standard Bayes prior can be seen as
the limit of a sequence of HA priors
where the
inverse-Wishart
prior distributions for the $\Sigma$-matrices converge to point masses
on the identity matrices of the appropriate dimension.

For each simulated data set,
the Gibbs sampler described in Section~\ref{sec:Model}
was run for 11,000 iterations,
the first 1000 of which were dropped to allow for convergence to the
stationary distribution. Parameter values
were saved every 10th scan, resulting in 1000 Monte Carlo samples per
simulation. Starting values for all the mean effects were set to zero
and all variances set to identity matrices of the proper dimensions.
We examined the convergence and autocorrelation of the marginal samples
of
the parameters in each procedure.
Since the number of parameters is large, we present
the results of Geweke's $z$-test and estimates of the
effective sample size for the error variance $\sigma^{2}$,
as it provides a parsimonious summary of the convergence results.
The minimum effective sample size across all simulations
was 233 out of the 1000 recorded scans, and the average
effective sample size was 895. Geweke's
$z$-statistic was less than $2$ in absolute value in 93, 93, 97 and 95
percent of the Markov
chains for the four sample sizes
(with the percentages being identical for both
Bayesian methods). While the cases in which $\vert z\vert>2$ were
not extensively examined, it is assumed that running the chain longer
would have yielded improved estimation.

\begin{figure}

\includegraphics{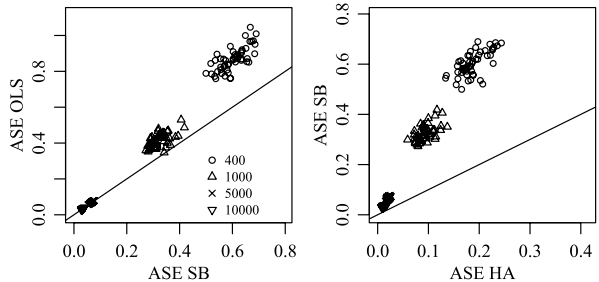}

\caption{Comparison of ASE for different estimation methods when the
true means array exhibits order consistent interactions.}
\label{Flo:mse_full}
\end{figure}

For each simulated data set, the posterior mean estimates $\hat
{M}_{\mathrm{HA}}$
and $\hat{M}_{\mathrm{SB}}$ were obtained by averaging
their values across the 1000 saved
iterations of the Gibbs sampler. The average squared
error (ASE) of estimation was calculated as $\operatorname{ASE}(\hat
{M})=\Vert\hat
{M}-M\Vert^{2}/(m_1 m_2 m_3)$,
where $M$ is the means array that generated the data. These values
were then compared across the three approaches. %
The left panel of Figure~\ref{Flo:mse_full} demonstrates that
the SB estimator provided a reduction in ASE when compared to the
OLS estimator for all data sets with sample sizes 400 and 1000, 96\%
of the data sets with sample size 5000 and 90\% of data sets with
sample size 10,000. The second panel demonstrates that the HA estimator
provides a substantial further reduction in ASE for all data sets.
As we would expect, the reduction in ASE is dependent on the sample
size and decreases as the sample size increases.

%

These results are not surprising:
By estimating the variances $\sigma_{a}^{2},\sigma_{ab}^{2}$, etc.
from the data,
the SB approach provides adaptive shrinkage
and so we expect these SB estimates to
outperform the
OLS estimates in terms of ASE.
However, the SB approach does not
use information on the similarity among the levels of an effect, and
so its estimation of higher order interactions relies on the limited
information available directly in the corresponding sufficient statistics.
As such, we expect the SB estimates to perform less well than the HA
estimates, which are able to borrow information
from well-estimated main effects and low-order interactions to assist
in the estimation of higher-order terms for which data information
is limited.

%
\begin{figure}

\includegraphics{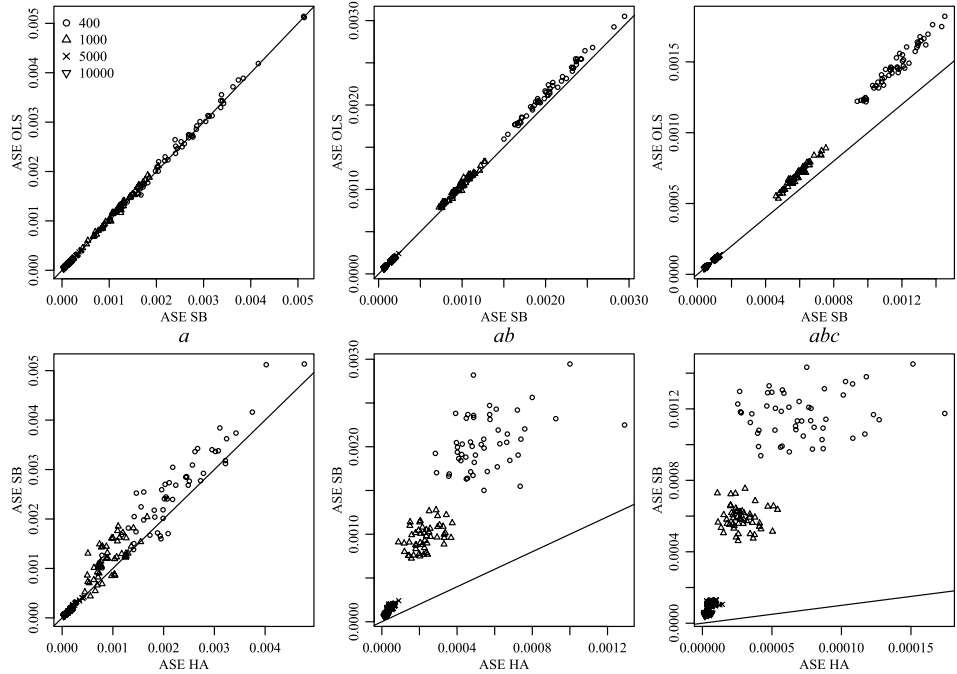}

\caption{ASE comparisons for the main effect, a two-way interaction and
a three-way interaction that involve $a$ are in the three columns,
respectively. The first row compares ASE between SB and OLS and the
second row compares ASE between HA and SB.}
\label{fig:full_ase_all}
\end{figure}

This behavior is further illustrated in Figure~\ref{fig:full_ase_all}
that provides an ASE comparison for the effects in the decomposition
of the means array.
To produce these plots, we decomposed each estimated means
array and considered the ASE for each effect when compared to the
decomposition of the true means array.%
It is immediate that the
gains in ASE are primarily from improved estimation of the higher
order interaction terms.
The top row of Figure~\ref{fig:full_ase_all} demonstrates that the SB
estimator performs
at least as well as the OLS estimator in terms of ASE for the main
effect $a$, and provides
a detectable reduction in ASE for two- and three-way interactions. The
reduction in ASE
for the higher order terms is due to the shrinkage provided by SB.
The second row of Figure~\ref{fig:full_ase_all} demonstrates that the
HA estimator provides a
moderate reduction in ASE for the main effect $a$ and a substantial
further reduction in ASE
for the higher order terms. This is exactly the behavior we expect, as
the HA procedure is able to borrow information from lower order terms
in order to further shrink higher order interactions.
We have also evaluated the width and coverage of nominal 95\% confidence
intervals for the cell means. The results for HA and SB are presented
in Table~\ref{tab:CI_OCI}.
The confidence intervals for the entries in the means 
array were smaller for the HA procedure than for SB, while the 
coverage was approximately 95\% for both. 
%
%
\begin{table}
\tablewidth=250pt
\caption{Actual coverage and interval widths of 95\% nominal confidence
intervals
for the cell means as estimated by HA and SB when order consistent
interactions are present}\label{tab:CI_OCI}
\begin{tabular*}{250pt}{@{\extracolsep{\fill}}lcccc@{}}
\hline
&\multicolumn{2}{c}{\textbf{Coverage}}&\multicolumn{2}{c@{}}{\textbf
{Width}}\\[-6pt]
&\multicolumn{2}{c}{\hrulefill}&\multicolumn{2}{c@{}}{\hrulefill}\\
\textbf{OBS} & \textbf{HA} & \textbf{SB} & \textbf{HA} & \textbf
{SB} \\
\hline
400 & 0.94 & 0.93 & 1.55 & 3.18 \\
1000 & 0.93 & 0.95 & 1.04 & 2.32 \\
5000 & 0.94 & 0.94 & 0.49 & 1.00 \\
10,000 & 0.95 & 0.95 & 0.36 & 0.70\\
\hline
\end{tabular*}
\end{table}

Recall that the parameters in the mean array $M$ were
generated by binning a third-degree polynomial, and were
not generated
from array normal distributions, that is, the HA prior is ``incorrect''
as a model for $M$. Even so, the HA prior is able to capture
the similarities between adjacent factor levels, resulting in
improved estimation.
However, we note that not all of the improvement in ASE achieved by the
HA prior should be attributed to the identification of order-consistent
interactions. The simulation study that follows suggests some of the
performance of the HA prior is due to additional parameter shrinkage
provided by the
inverse-Wishart distributions on the $\Sigma$-matrices.


\subsection{Data with order inconsistent interactions}
\label{sub:data-oii}

\begin{figure}

\includegraphics{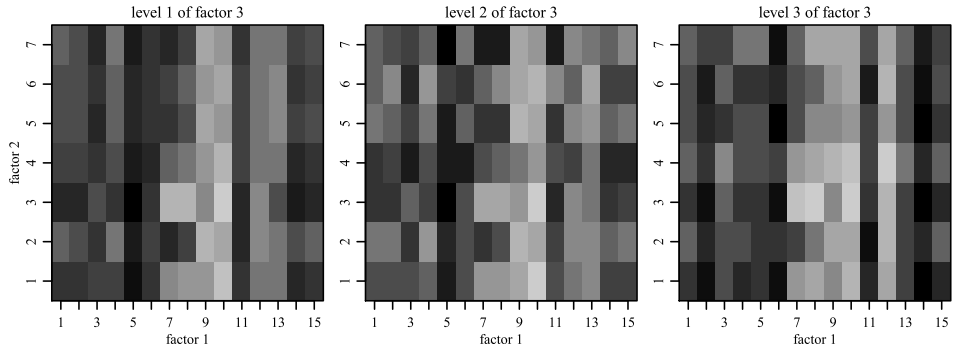}

\caption{The means array $M$ for the second simulation study, across
levels of the third factor.}
\label{Flo:marray_full_oii}
\end{figure}

In this subsection we evaluate the HA approach
for populations which exhibit interactions that are order inconsistent.
The means array $M$ is constructed by taking the means array from
Section~\ref{sub:Data-with-interactions}, decomposing it into main
effects, two- and three-way interactions, permuting the levels of each
factor within each effect, and reconstructing a means array.
That is, if $ \{a_{i}\dvtx i=1,\ldots,m_1 \}$ is the
collection of parameters for the first main effect and $ \{
(ab )_{ij}
\dvtx i=1,\ldots,m_1,j=1,\ldots,m_2 \}$ is the collection of parameters
for the
two-way interaction between the first and second factors in
Section~\ref{sub:Data-with-interactions}, then $ \{a_{\pi_1(i)} \}$ and
$ \{ (ab )_{\pi_2(i)\pi_3(j)} \}$ are the main effect
and two-way interaction parameters for the means array in this section,
where $\pi_1,\pi_2\mbox{ and }\pi_3$ are independent
permutations. The remaining effects were permuted analogously.
Due to this construction, the magnitudes of the main effects,
two- and three-way interactions remain the same, but the process
becomes less ``smooth,'' as can be seen in Figure~\ref{Flo:marray_full_oii}.

\begin{figure}[b]

\includegraphics{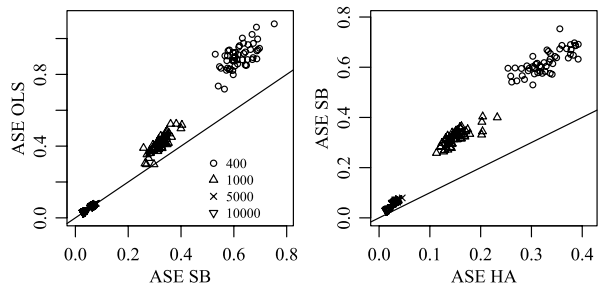}

\caption{Comparison of ASE for different estimation methods when the
true means array
exhibits order inconsistent interactions that have the same magnitude
as the order consistent interactions of Section \protect\ref
{sub:Data-with-interactions}.}
\label{Flo:mse_full_no_const}
\end{figure}

Again, 50 data sets were generated for each sample size, and estimates
$\hat{M}_\mathrm{HA}$, $\hat{M}_\mathrm{SB}$ and $\hat{M}_\mathrm
{OLS}$ were
obtained for each data set, where the Bayesian estimates were obtained
using the same MCMC approximation procedure as in the previous subsection.
Figure~\ref{Flo:mse_full_no_const} compares ASE across the different
approaches. The left panel of Figure~\ref{Flo:mse_full_no_const}, as
with the
left panel of Figure~\ref{Flo:mse_full}, demonstrates that the SB estimator
provides a reduction in ASE when compared to the OLS estimator. As expected,
since neither of these approaches take advantage of the structure of the
order consistent
interactions, this plot is nearly identical to the corresponding plot
in Figure~\ref{Flo:mse_full}.

%

The second panel demonstrates
that the HA estimator provides a further reduction in ASE for all data sets,
although this reduction is less substantial than in the presence
of order consistent interactions.
The lower ASE of the HA estimates may be initially surprising, as there
are no order consistent interactions for the HA prior to take advantage of.
We conjecture that the lower ASE is due to the additional shrinkage on the
parameter estimates that the inverse-Wishart priors on the
$\Sigma$-parameters provide. For example, under both the SB and HA priors
we have
$\operatorname{Cov}[\operatorname{vec}(ab)] = \Sigma_b \otimes
\Sigma_a /\gamma_{ab}$, but
under the former the covariance matrices are set to the identity,
whereas under the latter they have inverse-Wishart distributions.

%

As with the previous simulation, we evaluated the width
and coverage of nominal 95\% confidence
intervals for the cell means. The results for HA and SB are presented
in Table~\ref{tab:CI_OII}. As in the previous simulation, the coverage
for both procedures
is approximately 95\%. The confidence intervals
are wider for SB than for HA, but the differences
between the two procedures are much smaller in this
simulation as compared to the previous one.
\begin{table}
\tablewidth=250pt
\caption{Actual coverage and interval widths of 95\% nominal confidence
intervals
for the cell means as estimated by HA and SB when order inconsistent
interactions are present}\label{tab:CI_OII}
\begin{tabular*}{250pt}{@{\extracolsep{\fill}}lcccc@{}}
\hline
&\multicolumn{2}{c}{\textbf{Coverage}}&\multicolumn{2}{c@{}}{\textbf
{Width}}\\[-6pt]
&\multicolumn{2}{c}{\hrulefill}&\multicolumn{2}{c@{}}{\hrulefill}\\
\textbf{OBS} & \textbf{HA} & \textbf{SB} & \textbf{HA} & \textbf
{SB} \\
\hline
400 & 0.95 & 0.94 & 2.26 & 2.98 \\
1000 & 0.95 & 0.95 & 1.56 & 2.15 \\
5000 & 0.96 & 0.95 & 0.73 & 0.98 \\
10,000 & 0.96 & 0.94 & 0.53 & 0.69\\
\hline
\end{tabular*}
\end{table}

\subsection{Data with order inconsistent interactions: Bayes risk}

The surprising outcome of the previous section requires further
study of the behavior of the HA approach when order
inconsistent interactions are present. To get a more complete
picture of this behavior, we evaluate the Bayes risk of the procedure
when data is generated directly from the SB prior.
We construct 200 means arrays $M_1,\ldots,M_{200}$ of the same
dimensions as in
the previous subsections using the following procedure:
\begin{longlist}[1.]
\item[1.] Generate $\gamma_{a},\gamma_{b},\gamma_{c},\gamma
_{ab},\gamma
_{ac},\gamma_{bc},\gamma_{abc}\stackrel{\mathrm{i.i.d.}}{\sim
}\operatorname{gamma}
(\nu/2,\tau^2/2 )$
with shape paramter $\nu= 4$ and rate parameter $\tau^2=2$. These are
the precision components
for the 3 main effects, 3 two-way interactions and 1 three-way
interaction, respectively.
\item[2.] Generate effect levels as follows: $ \{a_1,\ldots,a_{15}
\}\sim N (0,I/\gamma_{a} )$, $ \{ab_{1,1},\ldots
,\break ab_{15,7} \}\sim N (0,I/\gamma_{ab} )$, and similarly
for the remaining 5 effects.
\item[3.] Combine the effects from (2) into a means array $M_i$ according
to equation~\eqref{eq:modelbalcomp}.
\end{longlist}
For each sample size $\{400,1000,5000,10\mbox{,}000\}$ we
generated 50
data sets, each using a unique means array $M_i$, in the same manner as
in the
previous two simulation studies.
We obtained estimates
${\widehat{M_{i}}}_{\mathrm{HA}}$, ${\widehat{M_{i}}}_{\mathrm
{SB}}$ and ${\widehat
{M_{i}}}_{\mathrm{OLS}}$
for each data set, where the Bayesian estimates were obtained
using the same MCMC procedure as in the previous two subsections.

ASE represents the posterior quadratic loss of an estimation procedure
for a particular data set, and so by
varying the true means array $M_i$ between simulated data sets,
we can estimate the Bayes risk of an estimation procedure by
taking the average of ASE across simulated data sets.
The Bayes risk for the SB procedure is guaranteed to be smaller
than that for OLS and HA for all sample sizes and so
we report the results of the simulation study as ratios of estimated
Bayes risk
for SB to the estimated Bayes risk of the other procedures in
Table~\ref{tab:bayes_risk}.
For example, the first entry in the top row of Table~\ref{tab:bayes_risk}
states that the Bayes risk for SB is 41\% lower than the Bayes
risk for the OLS procedure for a sample size of 400.
As is expected, the difference in Bayes risk shrinks with
increasing sample size for both OLS and HA. The results
of this simulation study suggest that even for moderately sized data sets,
the relative risk of using the HA procedure when compared to SB is
rather small even when all effects are completely independent.
Additionally, the posterior estimates of all of the effects in the
decomposition of the means array had similar variances under both
SB and HA priors.
This suggests that using the HA procedure is not detrimental even when the
``order consistency'' of the interactions cannot be verified.

%
\begin{table}
\tablewidth=250pt
\caption{Ratio of estimated Bayes risk for SB to OLS and HA by sample size}
\label{tab:bayes_risk}
\begin{tabular*}{250pt}{@{\extracolsep{\fill}}lcccc@{}}
\hline
\textbf{Sample size} & \textbf{400} & \textbf{1000} & \textbf{5000}
& \textbf{10\mbox{,}000} \\
\hline
OLS & 0.59 & 0.69 & 0.93 & 0.97 \\
HA & 0.78 & 0.91 & 0.97 & 0.98 \\
\hline
\end{tabular*}
\end{table}

\subsection{Data without interactions}\label{sub:Data-without-interactions}

In this subsection we evaluate the HA approach
for populations in which interactions are not present.
The data in this simulation is generated from a model where the means
array $M$ is exactly additive and was constructed by binning
a linear function of three variables. As in the previous simulations,
$M$ is of dimension $m_{1}\times m_{2}\times m_{3}=15\times7\times3$.
The magnitudes of the three
main effects are $\|a\|^2/m_1=3.0$, $\|b\|^2/m_2=1.3$ and $\|c\|^2/m_3=0.3$,
while all interactions are exactly zero.
In addition to the SB and OLS estimators, we compare the
HA approach
to two ``oracle'' estimators: the additive
model least squares
estimator (AOLS) and the Bayes estimator under the additive model (ASB).
The prior used by the ASB approach is the same as the SB prior,
but does not include
terms other than main effects in the model.

%
\begin{figure}

\includegraphics{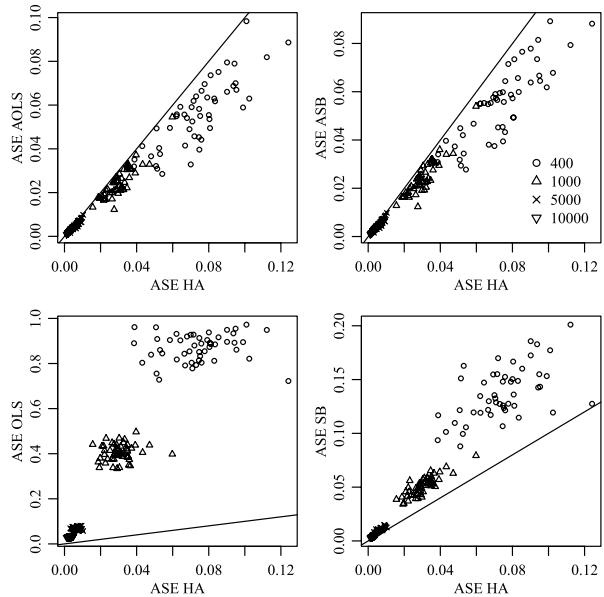}

\caption{Comparison of ASE for different estimation methods when the
true means array is additive.}
\label{Flo:mse_add}
\end{figure}

As before, 50 data sets were generated for each sample size, and estimates
$\hat{M}_{\mathrm{HA}}$, $\hat{M}_{\mathrm{SB}}$, $\hat
{M}_{\mathrm{OLS}}$,
$\hat{M}_{\mathrm{ASB}}$ and $\hat{M}_{\mathrm{AOLS}}$
were obtained
for each data set,
where the Bayesian estimates were obtained using the same
MCMC approximation procedure as in the previous two
subsections.
Some results are shown in
Figure~\ref{Flo:mse_add}, which
compares ASE across the different
approaches. In the top row of Figure~\ref{Flo:mse_add} we see that
the performance of the
HA estimates is comparable to but not as good as the
oracle least squares and Bayesian estimates in terms
of ASE.
Specifically, the ASE for the HA estimates is 24.2, 18.6, 20.1 and 17.4
percent higher than for the AOLS estimates for data sets with sample
sizes 400, 1000, 5000 and 10\mbox{,}000, respectively. Similarly, the
ASE for
the HA estimates is 25, 19.7, 20.3 and 17.8 percent higher than for the
ASB estimates for data sets with sample
sizes 400, 1000, 5000 and 10\mbox{,}000, respectively.
However,
the bottom
row of Figure~\ref{Flo:mse_add}
shows that the HA prior is superior to the other
nonoracle OLS and SB approaches
that attempt to estimate the interaction terms.

These results, together with those of the last two subsections,
suggest that the HA approach provides
a competitive method for fitting means arrays in the presence or
absence of
interactions. When order consistent interactions are present, the HA
approach is able to make use of the similarities across levels of the
factors, thereby outperforming approaches that cannot adapt to
such patterns.
Additionally, the
HA approach
does not appear to suffer when interactions are not order consistent.
Finally, in the absence of interactions altogether, the HA approach
adapts well, providing estimates similar to those that assume
the correct additive model.

%
%
%

\section{Analysis of carbohydrate intake}\label{sec:Data-analysis}
In this section we estimate average carbohydrate, sugar and fiber intake
by education, ethnicity and age using the HA procedure described in
Section~\ref{sec:Model}.
Our estimates are based on data from
2134 males from the US population, obtained from the 2007--2008
NHANES survey. Nutrient intake is self reported on two
nonconsecutive days.
Each day's data 
concerns
food and beverage intake from the preceding 24 hour period only, and is
calculated using
the USDA\textquoteright{}s Food and Nutrient Database for Dietary
Studies 4.1 [\citet{USDAfndds}]. All intake was measured in grams, and
we average the intake over
the two days to yield a single measurement per individual.
When intake information is only available for one day, we treat
that as the observation (we do not perform any reweighing to
account for this partial information). 
We are interested in relating the intake data to the following demographic
variables:
%
\begin{itemize}
\item Age: (31--40), (41--50), (51--60), (61--70), (71--80).
\item Education: Primary (P), Secondary (S), High School diploma (HD),
Associates
degree (AD), Bachelors degree (BD).
\item Ethnicity: Mexican (Hispanic), other Hispanic, white (not
Hispanic) and black
(not Hispanic). 
\end{itemize}
Sample sizes for age-education-ethnicity combination were presented in
Table~\ref{Flo:dat_sum} in Section~\ref{sec1}. Of the 2234 male respondents
within the above demographic groups,
100 were missing their nutrient intake information for both days,
with similar rates of missingness
across the demographic variables,
and were excluded from the analysis.
For the 2134 individuals included in the analysis,
291 were missing nutrient intake information one of the two days. For
those individuals, the available day's information was used as their
nutrient intake, while for the remaining 1843 individuals an
average over the two days was used.

\begin{figure}

\includegraphics{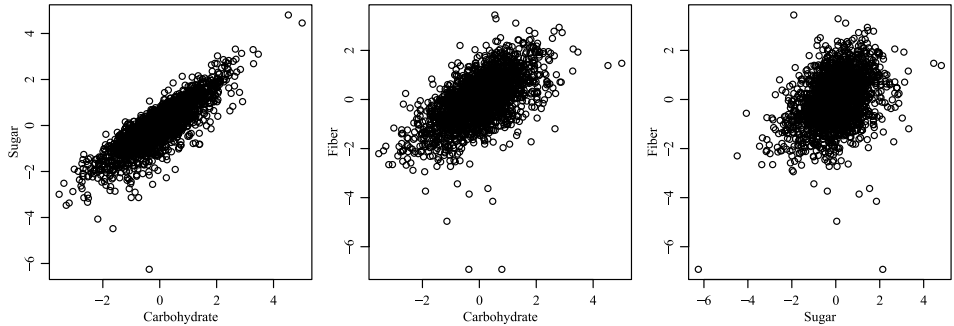}

\caption{Two-way plots of the transformed data.}
\label{Flo:trans-dat}
\end{figure}

The data on the original scale are somewhat skewed and show heteroscedasticity
across the demographic variables.
Since different variances across
groups can lead to bias in the sums of squares, making
$F$-tests for interactions anti-conservative
[\citet{miller1997beyond}],
stabilizing the variance is desirable.
Figure~\ref{Flo:trans-dat} provides two-way scatterplots of
the response variables
after applying a quarter power transformation to each variable, which we
found stabilized the variances across the groups better than either
a log or square-root transformation. Additionally, following the
quarter power
transformation, we centered
and scaled each response variable to have mean zero and variance one.

%

\subsection{MANOVA model and parameter estimation}\label{sec4.1}

As presented in
Table~\ref{Flo:manova-table} of Section~\ref{sec1},
$F$-tests indicate evidence for the presence of interactions in the
array of population cell means. However,
12\% of all age-education-ethnicity categories have sample sizes less
than 5,
and so we are concerned with overfitting of the OLS estimates.
As an alternative, we extend the HA methodology described in
Section~\ref{sec:Model}
to accommodate a MANOVA model. Our MANOVA model has the same form
as the ANOVA model given by equation (\ref{eq:modelbalcomp}), except that
each effect listed there is a three-dimensional vector corresponding
to the separate effects for each of the three response variables.
Additionally, the error terms now have a multivariate normal
distribution with zero-mean and unknown covariance matrix
$\Sigma_y$.


We extend the hierarchical array prior discussed above to accommodate
the $p$-variate MANOVA model as follows: Our prior for
the $m_1\times p$ matrix $a$ of main effects for the first
factor is
$\operatorname{vec}(a) \sim N_{m_1p}(0,I \otimes\Sigma_a)$, where
$\Sigma_a$ is as before.
Our prior
for the $m_1\times m_2\times p$ array $(ab)$ of two-way interaction
terms is given by
$\operatorname{vec}(ab) \sim N_{m_1m_2p}(0,
\Gamma_{ab}^{-1} \otimes\Sigma_b\otimes\Sigma_a)$.
Here, $\Gamma_{ab}$ is a $p\times p$ diagonal matrix whose terms
determine the scale of the two-way interactions for each of the
$p$ response variables.
If we consider only the first response, then $ (\Gamma_{ab}
)_{11}$ is
exactly the $\gamma_{ab}$ scalar described in the ANOVA setup.
Similarly, our prior for the four-way array $(abc)$ of three-way
interaction terms is
$\operatorname{vec}(abc) \sim N_{m_1m_2m_3p}(0,\Gamma
_{abc}^{-1}\otimes\Sigma_c
\otimes\Sigma_b\otimes\Sigma_a)$.
Priors for other main effects and interaction terms are defined similarly.
The hyperpriors for each diagonal entry of $\Gamma$ are independent
gamma distributions,
chosen as in Section~\ref{sec2.4} so that the prior magnitude of the
effects for
each response
is centered around the sum of squares of the effect from the OLS decomposition.

An alternative prior would be to include a
covariance matrix representing similarities of effects
across the three variables. This would be achieved by
replacing $I\otimes\Sigma_a$ in the prior for $a$ with
$\Sigma_p \otimes\Sigma_a$, $\Gamma_{ab}^{-1}$ with
$\Sigma_p \Gamma_{ab}^{-1}$ in the prior for $ab$, and so on.
Such a covariance term might be appropriate for data in which
marginal correlations between the $p$ response variables were
driven by similarities in the cell means, rather than by
within-cell correlations. In such a case we would
expect, for example, that
if $a_1$, the main effects for variable 1,
were positively correlated with $a_2$, the main effects for variable 2,
then $b_1$ and $b_2$ would be positively correlated, as
would $c_1$ and $c_2$, as well as any other pair
of effects in the decompositions of variables 1 and 2.
However, such consistency does not appear in our NHANES data:
For example, considering correlations between the ANOVA decomposition
parameters for
sugar and carbohydrates, we observe positive correlations for the main
effects of age and education and negative correlations for the
interaction terms age$\times$ethnicity and
age$\times$ethnicity$\times$education.
These observations support the choice of
$\Sigma_p=I$ in the prior for the analysis of
these data, although estimating $\Sigma_p$ might be warranted for
other data sets.

\begin{figure}

\includegraphics{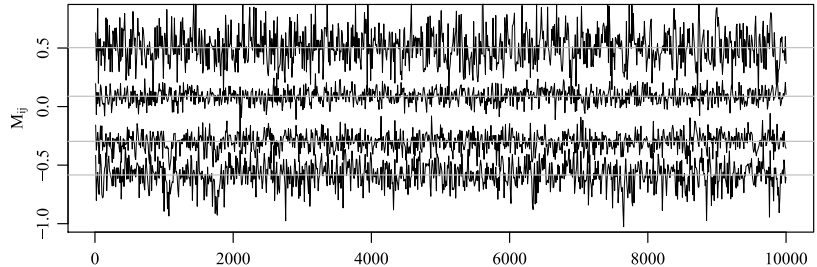}

\caption{MCMC samples of 4 out of 300 entries of the means
array $M$.}
\label{Flo:Mtrace}
\end{figure}

A Gibbs
sampling scheme similar to the one outlined in Section~\ref{sec:Model}
was iterated
200,000 times with parameter values saved every 10 scans, resulting
in 20,000 simulated values of the means array $M$ and the covariance
matrices $ \{ \Sigma_{\mathrm{eth}},\Sigma_{\mathrm{age}},\Sigma
_{\mathrm{edu}} \} $
for posterior analysis. Mixing of the Markov chain for $M$ was good:
Figure~\ref{Flo:Mtrace} shows MCMC samples of 4 out of 300 entries
of $M$ (chosen so
that their trace plots were visually distinct). The autocorrelation across
the saved scans was low, with the lag-10 autocorrelation for the
thinned chain less than 0.14 in absolute value for each element of
$M$ ($97.3\%$ of entries have lag-10 autocorrelation less than 0.07 in
absolute value) and effective sample sizes between 1929 and 13,520.
The mixing for the elements of the covariance matrices
$\Sigma_{\mathrm{eth}},\Sigma_{\mathrm{age}},\Sigma_{\mathrm{edu}}$
is not as good as that of the means array $M$: The maximum absolute
value of lag-10 autocorrelation
of the saved scans for the three rescaled covariance matrices is
0.18, 0.12 and 0.19, respectively. The effective sample sizes for the
elements of the
covariance matrices are at least 1684.%

%

%

\subsection{Posterior inference on $M$ and \texorpdfstring{$\Sigma$}{Sigma}s}\label{sec4.2}

\begin{figure}[t]
\includegraphics{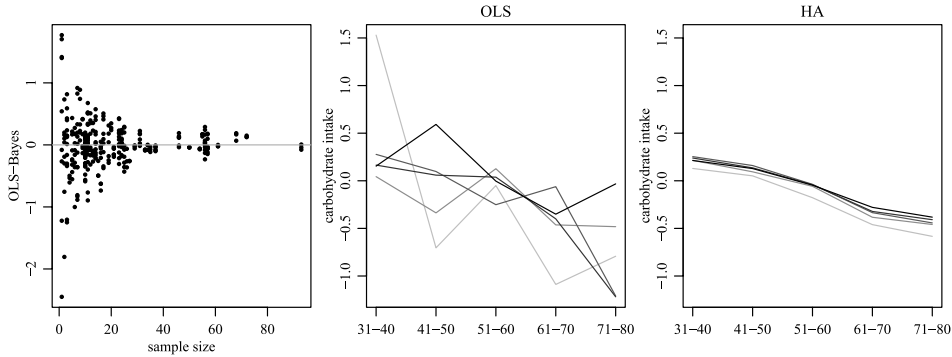}
\caption{Shrinkage and regularization plots. The first panel plots the
difference between the OLS
and HA estimates of a cell mean against the cell-specific sample sizes.
The second a third panels
plot estimated cell means of carbohydrate intake
for black survey participants across age and education levels, where
lighter shades represent
higher levels of education.}
\label{Flo:shr_reg}
\end{figure}

We obtain a Monte Carlo approximation to 
$\hat{M}=E [M|Y ]$ by averaging
over the saved scans of the Gibbs sampler.
Figure~\ref{Flo:shr_reg}
provides information on the shrinkage and regularization of the estimates
due to the HA procedure, as compared to OLS.
The first panel plots the difference between the OLS and Bayes
estimates of the
cell means versus cell-specific sample sizes.
For small sample sizes, the Bayes estimate for a given cell is affected
by the data from related cells, and can generally be quite different
from the OLS estimate (the cell sample mean). For cells with large sample
sizes the difference between the two estimates is generally small.
The second panel of the figure
plots the
OLS estimates of the cell means for carbohydrate intake of black survey
participants across
age and education levels.
Note that there appears to be a general trend of decreasing
intake with increasing age and education level, although the
OLS estimates themselves are not consistently ordered in this way.
In contrast, these trends are much more apparent in the Bayes estimates
plotted in the third panel.
The HA prior allows the parameter estimates to be close to additive,
while not enforcing strict additivity in this situation where we have
evidence of nonadditivity via the $F$-tests.
The smoothing provided by the HA prior is attributed
to its ability to share information across levels of an
effect and across interactions. When more levels are present
for a particular effect, the smoothing of the HA prior
closely resembles the behavior one would expect
from an unbinned continuous effect. On the other hand, OLS
will continue to model each cell-specific mean separately,
ignoring the similarities among levels and failing to
recognize the continuous nature of the effect.
The third panel of the figure was also more consistently ordered
than a similar analysis performed with the SB prior, suggesting
that the added shrinkage due to the inverse-Wishart priors and the
ability to share information across effect levels leads to
more realistic behavior of the estimates.

\begin{figure}

\includegraphics{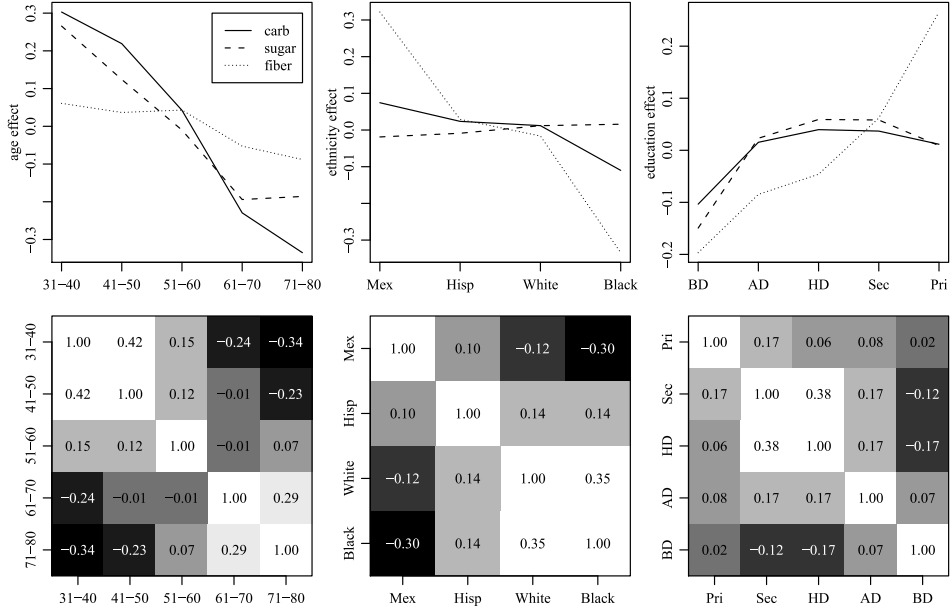}

\caption{Plots of main effects and interaction correlations for the three
outcome variables (carbohydrates, sugar and fiber). The first row of plots
gives HA estimates of the
main effects for each factor. The second row of plots
gives correlations of effects between levels of each factor, with
white representing 1 and darker colors representing a greater departure
from one.}
\label{Flo:post_effcors}
\end{figure}

The range of cell means for the centered and scaled effects
is $-0.58$ to 0.4 for carbohydrates,
$-0.38$ to 0.38 for sugar and $-1$ to 0.51 for fiber. The average standard
errors for the cell
means for the three responses are 0.08, 0.09 and 0.13, respectively.
When fitting
the data with the SB prior (analysis not included here), the average
standard errors
for the cell means were substantially larger: 0.12, 0.13 and 0.15 for
the three responses, respectively.
The first row of Figure~\ref{Flo:post_effcors} provides the estimates
of the main effects
from the HA procedure.
The second row of Figure~\ref{Flo:post_effcors}
summarizes covariance matrices $\{ \Sigma_\mathrm{eth},\Sigma
_\mathrm{age},\Sigma_\mathrm{edu}\}$
via the
posterior mean estimates of the correlation matrices
$ \{ C_{d,ij} \} = \{ \Sigma_{d,ij}/\sqrt{\Sigma
_{d,ii}\Sigma_{d,jj}} \} $
for $d\in\{ \mathrm{eth},\mathrm{age},\mathrm{edu} \} $.
In this figure, the
diagonal elements are all 1, and darker colors represent a greater
departure from one. The range of the estimated correlations was $-0.34$
to 0.42 for age categories, $-0.30$ to 0.35 for ethnic groups, and $-0.17$
to 0.38 for educational categories. For the two ordered categorical
variables, age and education, we see that closer categories are
generally more positively
correlated than ones that are further apart.
While the ethnicity variable is not ordered, its correlation matrix informs
us of which categories are more similar in terms of these response
variables. The middle panel of
the second row of Figure~\ref{Flo:post_effcors} confirms the
order-consistent interactions
we observed in Figure~\ref{Flo:two-way-manova}: Mexican
survey participants
are more similar to Hispanic participants in terms
of carbohydrate intake patterns than to white or black participants.

%

\begin{figure}

\includegraphics{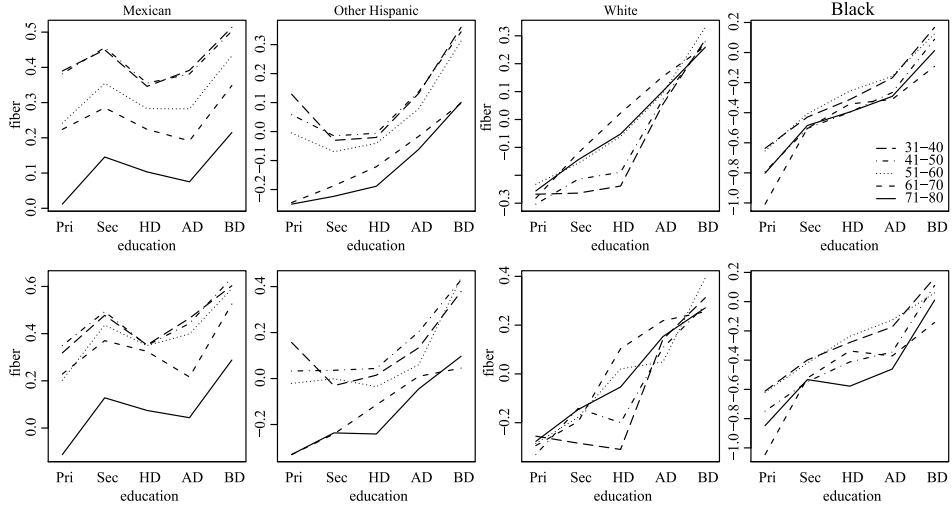}

\caption{HA and SB interaction plots of estimated mean fiber intake
by ethnicity, age and education level.
HA and SB estimates are in the top and bottom rows, respectively.}
\label{Flo:three-way-HA}
\end{figure}

For fiber intake, the top row of Figure~\ref{Flo:three-way-HA}
gives age by education interaction plots for each level of ethnicity,
using cell mean estimates obtained from the HA procedure.
Comparing these plots to the analogous plots for the
OLS estimates presented in Figure~\ref{Flo:three-way-OLS},
we see that the smoother HA estimates allow for a more
interpretable description of the three-way interaction.
Recall that a three-way interaction can be described as the
variability of a two-way interaction across levels of a
third factor.
Based on the plots,
the two-way age by education interactions
for the Mexican and Black groups seem quite small.
In contrast, the White and Other Hispanic groups
appear to have interactions that can be described as heterogeneity
in the education effect across levels of age.
For both of these groups, this heterogeneity is ordered by age:
For the Other Hispanic group, the education effects seem similar
for the three youngest age groups. For the White group, the
education effects seem similar for the two youngest age groups.

This similarity in education effects for similar levels of age
is more apparent in these HA estimates than in the corresponding
parameter estimates from the SB procedure, presented in the
second row of Figure~\ref{Flo:three-way-HA}, particularly for the White
ethnicity. In contrast to the SB approach,
the HA procedure was able to recognize the
similarity of parameters corresponding to adjacent age levels
and to use this information to assist with estimation.
Our expectations that age effects are likely to
be smooth, as well as the good performance of the HA procedure
in the simulation study of the previous section,
give us confidence that the HA procedure is providing more
realistic and interpretable cell mean estimates than
either the OLS or SB approaches.

%


\section{Discussion}\label{sec5}

This article has presented a novel hierarchical Bayes method for parameter
estimation of cross-classified data under ANOVA and MANOVA models.
Unlike least-squares
estimation, a Bayesian approach provides for regularized estimates of
the potentially large number of parameters in a MANOVA model. Unlike
the nonhierarchical Bayesian approach, the hierarchical approach
provides a data-driven method of regularization, and unlike the standard
hierarchical Bayes, the hierarchical array prior can identify
similarities among categories and
share this information across interaction effects to assist in the
estimation of higher-order terms for which data information is limited.
In a simulation study the HA approach was able to detect interactions
when they were present, and to estimate the means array better than
a full least squares or standard Bayesian approaches (in terms of
mean squared error). When the true means array was completely additive,
the HA prior was able to adapt to this smaller model better than the other
full model estimation approaches under consideration.

An immediate extension to our approach modifies the priors
on the covariance matrices to incorporate known
structure. For example, in the case of observations
for different time periods, an autoregressive
covariance model might be desirable. In the simplest case
of an $\operatorname{AR}(1)$ model, \citet{berger1994noninformative}
suggest the use of a reference prior $\pi_R (\rho)$ for the single
parameter $\rho$. We also note that due to the
scale nonidentifiability of the Kronecker product
we can assume that the variance parameter is equal to 1.
The posterior approximation follows the outline of Section~\ref
{sub:fullcond}: the full conditionals for the effects
and the full conditionals for the covariance
matrices that do not exhibit a specific structure remain the same.
The only difference is in the posterior approximation procedure
for the structured covariance matrix, where a posterior sample
of $\rho$ can be obtained by importance sampling.
The HA procedure can easily accommodate other
structured covariances as well, with the only changes to the
posterior approximation steps reflecting this additional prior
information for the covariance matrix.

Generalizations of the HA prior are applicable to
any model whose parameters consist of vectors, matrices and arrays
for which some of the index sets are shared.
This includes generalized linear models with categorical factors,
as well as ANCOVA models that involve interactions
between continuous and categorical explanatory variables.
As an example of the latter case,
suppose we are interested in estimating the
linear relationship between an outcome and a set of
explanatory variables for every combination of three categorical
factors.
The regression parameters then consist of an $m_1\times m_2\times
m_3\times p$
array, where $m_1,m_2,m_3$ are the numbers of factor levels and
$p$ is the number of continuous regressors.
The usual ANCOVA decomposition can be used to parametrize this
array in terms of main effects and interactions arrays,
for which a hierarchical
array prior may be used.

Computer code and data for the results in Sections~\ref
{sec:Simulation-study} and \ref{sec:Data-analysis} are available
in the supplementary material [\citet{volf2013code}].

\begin{supplement}[id=suppA]
\stitle{Data and code for simulations and analysis}
\slink[doi,text={10.1214/13-AOAS685\break SUPP}]{10.1214/13-AOAS685SUPP} 
\sdatatype{.zip}
\sfilename{aoas685\_supp.zip}
\sdescription{A bundle containing data sets and code files to perform
the simulations and data analysis.}
\end{supplement}

%


\printaddresses


\begin{thebibliography}{35}

\bibitem[\protect\citeauthoryear{Albrink and
Ullrich}{1986}]{albrink1986interaction}
%
\begin{barticle}[pbm]
\bauthor{\bsnm{Albrink},~\bfnm{M.~J.}\binits{M.~J.}} \AND
\bauthor{\bsnm{Ullrich},~\bfnm{I.~H.}\binits{I.~H.}}
(\byear{1986}).
\btitle{Interaction of dietary sucrose and fiber on serum lipids in healthy
young men fed high carbohydrate diets}.
\bjournal{Am. J. Clin. Nutr.}
\bvolume{43}
\bpages{419--428}.
\bid{issn={0002-9165}, pmid={3006471}}
\bptok{imsref}%
\end{barticle}
%
\endbibitem

\bibitem[\protect\citeauthoryear{Austin, Ogden and
Hill}{2011}]{austin2011trends}
%
\begin{barticle}[pbm]
\bauthor{\bsnm{Austin},~\bfnm{Gregory~L.}\binits{G.~L.}},
\bauthor{\bsnm{Ogden},~\bfnm{Lorraine~G.}\binits{L.~G.}} \AND
\bauthor{\bsnm{Hill},~\bfnm{James~O.}\binits{J.~O.}}
(\byear{2011}).
\btitle{Trends in carbohydrate, fat, and protein intakes and
association with
energy intake in normal-weight, overweight, and obese individuals:
1971--2006}.
\bjournal{Am. J. Clin. Nutr.}
\bvolume{93}
\bpages{836--843}.
\bid{doi={10.3945/ajcn.110.000141}, issn={1938-3207}, pii={ajcn.110.000141},
pmid={21310830}}
\bptok{imsref}%
\end{barticle}
%
\endbibitem

\bibitem[\protect\citeauthoryear{Basiotis et~al.}{1989}]{basiotis1989sources}
%
\begin{barticle}[pbm]
\bauthor{\bsnm{Basiotis},~\bfnm{P.~P.}\binits{P.~P.}},
\bauthor{\bsnm{Thomas},~\bfnm{R.~G.}\binits{R.~G.}},
\bauthor{\bsnm{Kelsay},~\bfnm{J.~L.}\binits{J.~L.}} \AND
\bauthor{\bsnm{Mertz},~\bfnm{W.}\binits{W.}}
(\byear{1989}).
\btitle{Sources of variation in energy intake by men and women as determined
from one year's daily dietary records}.
\bjournal{Am. J. Clin. Nutr.}
\bvolume{50}
\bpages{448--453}.
\bid{issn={0002-9165}, pmid={2773824}}
\bptok{imsref}%
\end{barticle}
%
\endbibitem

\bibitem[\protect\citeauthoryear{Beran}{2005}]{beran2005asp}
%
\begin{barticle}[mr]
\bauthor{\bsnm{Beran},~\bfnm{Rudolf}\binits{R.}}
(\byear{2005}).
\btitle{A{SP} fits to multi-way layouts}.
\bjournal{Ann. Inst. Statist. Math.}
\bvolume{57}
\bpages{201--220}.
\bid{doi={10.1007/BF02507022}, issn={0020-3157}, mr={2160647}}
\bptok{imsref}%
\end{barticle}
%
\endbibitem

\bibitem[\protect\citeauthoryear{Berger and
Yang}{1994}]{berger1994noninformative}
%
\begin{barticle}[mr]
\bauthor{\bsnm{Berger},~\bfnm{James~O.}\binits{J.~O.}} \AND
\bauthor{\bsnm{Yang},~\bfnm{Ruo-yong}\binits{R.-y.}}
(\byear{1994}).
\btitle{Noninformative priors and {B}ayesian testing for the
{$\operatorname{AR}(1)$}
model}.
\bjournal{Econometric Theory}
\bvolume{10}
\bpages{461--482}.
\bid{doi={10.1017/S026646660000863X}, issn={0266-4666}, mr={1309107}}
\bptok{imsref}%
\end{barticle}
%
\endbibitem

\bibitem[\protect\citeauthoryear{Chandalia
et~al.}{2000}]{chandalia2000beneficial}
%
\begin{barticle}[pbm]
\bauthor{\bsnm{Chandalia},~\bfnm{M.}\binits{M.}},
\bauthor{\bsnm{Garg},~\bfnm{A.}\binits{A.}},
\bauthor{\bsnm{Lutjohann},~\bfnm{D.}\binits{D.}}, \bauthor
{\bparticle{von}
\bsnm{Bergmann},~\bfnm{K.}\binits{K.}},
\bauthor{\bsnm{Grundy},~\bfnm{S.~M.}\binits{S.~M.}} \AND
\bauthor{\bsnm{Brinkley},~\bfnm{L.~J.}\binits{L.~J.}}
(\byear{2000}).
\btitle{Beneficial effects of high dietary fiber intake in patients
with type 2
diabetes mellitus}.
\bjournal{N. Engl. J. Med.}
\bvolume{342}
\bpages{1392--1398}.
\bid{doi={10.1056/NEJM200005113421903}, issn={0028-4793}, pmid={10805824}}
\bptok{imsref}%
\end{barticle}
%
\endbibitem

\bibitem[\protect\citeauthoryear{Cui et~al.}{2010}]{cui2010partitioning}
%
\begin{barticle}[mr]
\bauthor{\bsnm{Cui},~\bfnm{Yue}\binits{Y.}},
\bauthor{\bsnm{Hodges},~\bfnm{James~S.}\binits{J.~S.}},
\bauthor{\bsnm{Kong},~\bfnm{Xiaoxiao}\binits{X.}} \AND
\bauthor{\bsnm{Carlin},~\bfnm{Bradley~P.}\binits{B.~P.}}
(\byear{2010}).
\btitle{Partitioning degrees of freedom in hierarchical and other richly
parameterized models}.
\bjournal{Technometrics}
\bvolume{52}
\bpages{124--136}.
\bid{doi={10.1198/TECH.2009.08161}, issn={0040-1706}, mr={2752111}}
\bptok{imsref}%
\end{barticle}
%
\endbibitem

\bibitem[\protect\citeauthoryear{Dawid}{1981}]{dawid1981some}
%
\begin{barticle}[mr]
\bauthor{\bsnm{Dawid},~\bfnm{A.~P.}\binits{A.~P.}}
(\byear{1981}).
\btitle{Some matrix-variate distribution theory: Notational
considerations and
a {B}ayesian application}.
\bjournal{Biometrika}
\bvolume{68}
\bpages{265--274}.
\bid{doi={10.1093/biomet/68.1.265}, issn={0006-3444}, mr={0614963}}
\bptok{imsref}%
\end{barticle}
%
\endbibitem

\bibitem[\protect\citeauthoryear{Friedman, Hastie and
Tibshirani}{2010}]{friedman2010note}
%
\begin{bmisc}[auto:STB|2013/10/14|10:36:11]
\bauthor{\bsnm{Friedman},~\bfnm{J.}\binits{J.}},
\bauthor{\bsnm{Hastie},~\bfnm{T.}\binits{T.}} \AND
\bauthor{\bsnm{Tibshirani},~\bfnm{R.}\binits{R.}}
(\byear{2010}).
\bhowpublished{A note on the group lasso and a sparse group lasso.
Available at \arxivurl{arXiv:1001.0736}.}
\bptok{imsref}%
\end{bmisc}
%
\endbibitem

\bibitem[\protect\citeauthoryear{Gelman}{2005}]{gelman2005analysis}
%
\begin{barticle}[mr]
\bauthor{\bsnm{Gelman},~\bfnm{Andrew}\binits{A.}}
(\byear{2005}).
\btitle{Analysis of variance---Why it is more important than ever}.
\bjournal{Ann. Statist.}
\bvolume{33}
\bpages{1--53}.
\bid{doi={10.1214/009053604000001048}, issn={0090-5364}, mr={2157795}}
\bptnote{check related}%
\bptok{imsref}%
\end{barticle}
%
\endbibitem

\bibitem[\protect\citeauthoryear{Gelman and Hill}{2007}]{gelman2007data}
%
\begin{bmisc}[auto:STB|2013/10/14|10:36:11]
\bauthor{\bsnm{Gelman},~\bfnm{A.}\binits{A.}} \AND
\bauthor{\bsnm{Hill},~\bfnm{J.}\binits{J.}}
(\byear{2007}).
\bhowpublished{Data analysis using regression and multilevel hierarchical
models. Unpublished manuscript}.
\bptok{imsref}%
\end{bmisc}
%
\endbibitem\vadjust{\goodbreak}

\bibitem[\protect\citeauthoryear{Genkin, Lewis and
Madigan}{2007}]{genkin2007large}
%
\begin{barticle}[mr]
\bauthor{\bsnm{Genkin},~\bfnm{Alexander}\binits{A.}},
\bauthor{\bsnm{Lewis},~\bfnm{David~D.}\binits{D.~D.}} \AND
\bauthor{\bsnm{Madigan},~\bfnm{David}\binits{D.}}
(\byear{2007}).
\btitle{Large-scale {B}ayesian logistic regression for text categorization}.
\bjournal{Technometrics}
\bvolume{49}
\bpages{291--304}.
\bid{doi={10.1198/004017007000000245}, issn={0040-1706}, mr={2408634}}
\bptok{imsref}%
\end{barticle}
%
\endbibitem

\bibitem[\protect\citeauthoryear{Hodges et~al.}{2007}]{hodges2007smoothing}
%
\begin{barticle}[mr]
\bauthor{\bsnm{Hodges},~\bfnm{James~S.}\binits{J.~S.}},
\bauthor{\bsnm{Sargent},~\bfnm{Daniel~J.}\binits{D.~J.}},
\bauthor{\bsnm{Cui},~\bfnm{Yue}\binits{Y.}} \AND
\bauthor{\bsnm{Carlin},~\bfnm{Bradley~P.}\binits{B.~P.}}
(\byear{2007}).
\btitle{Smoothing balanced single-error-term analysis of variance}.
\bjournal{Technometrics}
\bvolume{49}
\bpages{12--25}.
\bid{doi={10.1198/004017006000000408}, issn={0040-1706}, mr={2345448}}
\bptok{imsref}%
\end{barticle}
%
\endbibitem

\bibitem[\protect\citeauthoryear{Hoff}{2011}]{hoff2011separable}
%
\begin{barticle}[mr]
\bauthor{\bsnm{Hoff},~\bfnm{Peter~D.}\binits{P.~D.}}
(\byear{2011}).
\btitle{Separable covariance arrays via the {T}ucker product, with applications
to multivariate relational data}.
\bjournal{Bayesian Anal.}
\bvolume{6}
\bpages{179--196}.
\bid{doi={10.1214/11-BA606}, issn={1936-0975}, mr={2806238}}
\bptok{imsref}%
\end{barticle}
%
\endbibitem

\bibitem[\protect\citeauthoryear{Johansson
et~al.}{2001}]{johansson2001underreporting}
%
\begin{barticle}[auto:STB|2013/10/14|10:36:11]
\bauthor{\bsnm{Johansson},~\bfnm{G.}\binits{G.}},
\bauthor{\bsnm{Wikman},~\bfnm{A.}\binits{A.}},
\bauthor{\bsnm{Ahren},~\bfnm{A.~M.}\binits{A.~M.}},
\bauthor{\bsnm{Hallmans},~\bfnm{G.}\binits{G.}} \AND
\bauthor{\bsnm{Johansson},~\bfnm{I.}\binits{I.}} \betal{et al.}
(\byear{2001}).
\btitle{Underreporting of energy intake in repeated 24-hour recalls
related to
gender, age, weight status, day of interview, educational level, reported
food intake, smoking habits and area of living}.
\bjournal{Public Health Nutrition}
\bvolume{4}
\bpages{919--928}.
\bptok{imsref}%
\end{barticle}
%
\endbibitem

\bibitem[\protect\citeauthoryear{Kass and
Wasserman}{1995}]{kasswasserman1995}
%
\begin{barticle}[mr]
\bauthor{\bsnm{Kass},~\bfnm{Robert~E.}\binits{R.~E.}} \AND
\bauthor{\bsnm{Wasserman},~\bfnm{Larry}\binits{L.}}
(\byear{1995}).
\btitle{A reference {B}ayesian test for nested hypotheses and its relationship
to the {S}chwarz criterion}.
\bjournal{J. Amer. Statist. Assoc.}
\bvolume{90}
\bpages{928--934}.
\bid{issn={0162-1459}, mr={1354008}}
\bptok{imsref}%
\end{barticle}
%
\endbibitem

\bibitem[\protect\citeauthoryear{Kolda and Bader}{2009}]{kolda2009tensor}
%
\begin{barticle}[mr]
\bauthor{\bsnm{Kolda},~\bfnm{Tamara~G.}\binits{T.~G.}} \AND
\bauthor{\bsnm{Bader},~\bfnm{Brett~W.}\binits{B.~W.}}
(\byear{2009}).
\btitle{Tensor decompositions and applications}.
\bjournal{SIAM Rev.}
\bvolume{51}
\bpages{455--500}.
\bid{doi={10.1137/07070111X}, issn={0036-1445}, mr={2535056}}
\bptok{imsref}%
\end{barticle}
%
\endbibitem

\bibitem[\protect\citeauthoryear{Kruschke}{2011}]{kruschke2010doing}
%
\begin{bbook}[auto:STB|2013/10/14|10:36:11]
\bauthor{\bsnm{Kruschke},~\bfnm{J.}\binits{J.}}
(\byear{2011}).
\btitle{Doing Bayesian Data Analysis: A Tutorial Introduction with R
and BUGS}.
\bpublisher{Academic Press}, \blocation{Boston, MA}.
\bptok{imsref}%
\end{bbook}
%
\endbibitem

\bibitem[\protect\citeauthoryear{Miller and Brown}{1997}]{miller1997beyond}
%
\begin{bbook}[auto:STB|2013/10/14|10:36:11]
\bauthor{\bsnm{Miller},~\bfnm{R.}\binits{R.}} \AND
\bauthor{\bsnm{Brown},~\bfnm{B.}\binits{B.}}
(\byear{1997}).
\btitle{Beyond ANOVA: Basics of Applied Statistics}.
\bpublisher{Chapman \& Hall/CRC}, \blocation{New York}.
\bptok{imsref}%
\end{bbook}
%
\endbibitem

\bibitem[\protect\citeauthoryear{Moerman, De~Mesquita and
Runia}{1993}]{moerman1993dietary}
%
\begin{barticle}[auto:STB|2013/10/14|10:36:11]
\bauthor{\bsnm{Moerman},~\bfnm{C.}\binits{C.}},
\bauthor{\bsnm{De~Mesquita},~\bfnm{H.}\binits{H.}} \AND
\bauthor{\bsnm{Runia},~\bfnm{S.}\binits{S.}}
(\byear{1993}).
\btitle{Dietary sugar intake in the aetiology of biliary tract cancer}.
\bjournal{International Journal of Epidemiology}
\bvolume{22}
\bpages{207--214}.
\bptok{imsref}%
\end{barticle}
%
\endbibitem

\bibitem[\protect\citeauthoryear{Montonen et~al.}{2003}]{montonen2003whole}
%
\begin{barticle}[pbm]
\bauthor{\bsnm{Montonen},~\bfnm{Jukka}\binits{J.}},
\bauthor{\bsnm{Knekt},~\bfnm{Paul}\binits{P.}},
\bauthor{\bsnm{J{\"{a}}rvinen},~\bfnm{Ritva}\binits{R.}},
\bauthor{\bsnm{Aromaa},~\bfnm{Arpo}\binits{A.}} \AND
\bauthor{\bsnm{Reunanen},~\bfnm{Antti}\binits{A.}}
(\byear{2003}).
\btitle{Whole-grain and fiber intake and the incidence of type 2 diabetes}.
\bjournal{Am. J. Clin. Nutr.}
\bvolume{77}
\bpages{622--629}.
\bid{issn={0002-9165}, pmid={12600852}}
\bptok{imsref}%
\end{barticle}
%
\endbibitem

\bibitem[\protect\citeauthoryear{Nielsen and
Popkin}{2004}]{nielsen2004changes}
%
\begin{barticle}[pbm]
\bauthor{\bsnm{Nielsen},~\bfnm{Samara~Joy}\binits{S.~J.}} \AND
\bauthor{\bsnm{Popkin},~\bfnm{Barry~M.}\binits{B.~M.}}
(\byear{2004}).
\btitle{Changes in beverage intake between 1977 and 2001}.
\bjournal{Am. J. Prev. Med.}
\bvolume{27}
\bpages{205--210}.
\bid{doi={10.1016/j.amepre.2004.05.005}, issn={0749-3797},
pii={S0749-3797(04)00122-9}, pmid={15450632}}
\bptok{imsref}%
\end{barticle}
%
\endbibitem

\bibitem[\protect\citeauthoryear{Olson}{1976}]{olson1976choosing}
%
\begin{barticle}[auto:STB|2013/10/14|10:36:11]
\bauthor{\bsnm{Olson},~\bfnm{C.}\binits{C.}}
(\byear{1976}).
\btitle{On choosing a test statistic in multivariate analysis of variance}.
\bjournal{Psychological Bulletin}
\bvolume{83}
\bpages{579}.
\bptok{imsref}%
\end{barticle}
%
\endbibitem

\bibitem[\protect\citeauthoryear{Park and Casella}{2008}]{park2008bayesian}
%
\begin{barticle}[mr]
\bauthor{\bsnm{Park},~\bfnm{Trevor}\binits{T.}} \AND
\bauthor{\bsnm{Casella},~\bfnm{George}\binits{G.}}
(\byear{2008}).
\btitle{The {B}ayesian lasso}.
\bjournal{J. Amer. Statist. Assoc.}
\bvolume{103}
\bpages{681--686}.
\bid{doi={10.1198/016214508000000337}, issn={0162-1459}, mr={2524001}}
\bptok{imsref}%
\end{barticle}
%
\endbibitem

\bibitem[\protect\citeauthoryear{Park, Gelman and
Bafumi}{2006}]{park2006state}
%
\begin{bmisc}[auto:STB|2013/10/14|10:36:11]
\bauthor{\bsnm{Park},~\bfnm{D.}\binits{D.}},
\bauthor{\bsnm{Gelman},~\bfnm{A.}\binits{A.}} \AND
\bauthor{\bsnm{Bafumi},~\bfnm{J.}\binits{J.}}
(\byear{2006}).
\bhowpublished{State level opinions from national surveys: Poststratification
using multilevel logistic regression. In \textit{Public Opinion in State
Politics} 209--228. Stanford Univ. Press, Stanford, CA}.
\bptok{imsref}%
\end{bmisc}
%
\endbibitem

\bibitem[\protect\citeauthoryear{Park et~al.}{2011}]{park2011dietary}
%
\begin{barticle}[pbm]
\bauthor{\bsnm{Park},~\bfnm{Yikyung}\binits{Y.}},
\bauthor{\bsnm{Subar},~\bfnm{Amy~F.}\binits{A.~F.}},
\bauthor{\bsnm{Hollenbeck},~\bfnm{Albert}\binits{A.}} \AND
\bauthor{\bsnm{Schatzkin},~\bfnm{Arthur}\binits{A.}}
(\byear{2011}).
\btitle{Dietary fiber intake and mortality in the NIH-AARP diet and health
study}.
\bjournal{Arch. Intern. Med.}
\bvolume{171}
\bpages{1061--1068}.
\bid{doi={10.1001/archinternmed.2011.18}, issn={1538-3679}, mid={NIHMS414391},
pii={archinternmed.2011.18}, pmcid={3513325}, pmid={21321288}}
\bptok{imsref}%
\end{barticle}
%
\endbibitem

\bibitem[\protect\citeauthoryear{Pittau, Zelli and
Gelman}{2010}]{pittau2010economic}
%
\begin{barticle}[auto:STB|2013/10/14|10:36:11]
\bauthor{\bsnm{Pittau},~\bfnm{M.}\binits{M.}},
\bauthor{\bsnm{Zelli},~\bfnm{R.}\binits{R.}} \AND
\bauthor{\bsnm{Gelman},~\bfnm{A.}\binits{A.}}
(\byear{2010}).
\btitle{Economic disparities and life satisfaction in European regions}.
\bjournal{Social Indicators Research}
\bvolume{96}
\bpages{339--361}.
\bptok{imsref}%
\end{barticle}
%
\endbibitem

\bibitem[\protect\citeauthoryear{USDA}{2010}]{USDAfndds}
%
\begin{bmisc}[auto:STB|2013/10/14|10:36:11]
\borganization{USDA}.
(\byear{2010}).
\bhowpublished{Food and Nutrient Database for Dietary Studies 4.1. U.S.
Dept. Agriculture, Agricultural Research Service, Food Surveys
Research Group, Beltsville, MD}.
\bptok{imsref}%
\end{bmisc}
%
\endbibitem

\bibitem[\protect\citeauthoryear{Verly~Junior
et~al.}{2010}]{verly2010sources}
%
\begin{barticle}[auto:STB|2013/10/14|10:36:11]
\bauthor{\bsnm{Verly~Junior},~\bfnm{E.}\binits{E.}},
\bauthor{\bsnm{Fisberg},~\bfnm{R.~M.}\binits{R.~M.}},
\bauthor{\bsnm{Cesar},~\bfnm{C.~L.~G.}\binits{C.~L.~G.}} \AND
\bauthor{\bsnm{Marchioni},~\bfnm{D.~M.~L.}\binits{D.~M.~L.}}
(\byear{2010}).
\btitle{Sources of variation of energy and nutrient intake among
adolescents in
S\~ao Paulo}.
\bjournal{Brazil. \textit{Cadernos de Sa\'ude P\'ublica}}
\bvolume{26}
\bpages{2129--2137}.
\bptok{imsref}%
\end{barticle}
%
\endbibitem

\bibitem[\protect\citeauthoryear{Volfovsky and Hoff}{2013}]{volf2013code}
%
\begin{bmisc}[auto:STB|2013/10/14|10:36:11]
\bauthor{\bsnm{Volfovsky},~\bfnm{A.}\binits{A.}} \AND
\bauthor{\bsnm{Hoff},~\bfnm{P.}\binits{P.}}
(\byear{2013}).
\bhowpublished{Supplement to ``Hierarchical array priors for ANOVA
decompositions of cross-classified data.''
DOI:\doiurl{10.1214/13-AOAS685SUPP}.}
\bptok{imsref}%
\end{bmisc}
%
\endbibitem

\bibitem[\protect\citeauthoryear{Yang et~al.}{2003}]{yang2003carbohydrate}
%
\begin{barticle}[auto:STB|2013/10/14|10:36:11]
\bauthor{\bsnm{Yang},~\bfnm{E.~J.}\binits{E.~J.}},
\bauthor{\bsnm{Chung},~\bfnm{H.~K.}\binits{H.~K.}},
\bauthor{\bsnm{Kim},~\bfnm{W.~Y.}\binits{W.~Y.}},
\bauthor{\bsnm{Kerver},~\bfnm{J.~M.}\binits{J.~M.}} \AND
\bauthor{\bsnm{Song},~\bfnm{W.~O.}\binits{W.~O.}}
(\byear{2003}).
\btitle{Carbohydrate intake is associated with diet quality and risk factors
for cardiovascular disease in us adults: Nhanes iii}.
\bjournal{Journal of the American College of Nutrition}
\bvolume{22}
\bpages{71--79}.
\bptok{imsref}%
\end{barticle}
%
\endbibitem

\bibitem[\protect\citeauthoryear{Yuan and Lin}{2005}]{yuan2005efficient}
%
\begin{barticle}[mr]
\bauthor{\bsnm{Yuan},~\bfnm{Ming}\binits{M.}} \AND
\bauthor{\bsnm{Lin},~\bfnm{Yi}\binits{Y.}}
(\byear{2005}).
\btitle{Efficient empirical {B}ayes variable selection and estimation
in linear
models}.
\bjournal{J. Amer. Statist. Assoc.}
\bvolume{100}
\bpages{1215--1225}.
\bid{doi={10.1198/016214505000000367}, issn={0162-1459}, mr={2236436}}
\bptok{imsref}%
\end{barticle}
%
\endbibitem

\bibitem[\protect\citeauthoryear{Yuan and Lin}{2006}]{yuan2006model}
%
\begin{barticle}[mr]
\bauthor{\bsnm{Yuan},~\bfnm{Ming}\binits{M.}} \AND
\bauthor{\bsnm{Lin},~\bfnm{Yi}\binits{Y.}}
(\byear{2006}).
\btitle{Model selection and estimation in regression with grouped variables}.
\bjournal{J. R. Stat. Soc. Ser. B Stat. Methodol.}
\bvolume{68}
\bpages{49--67}.
\bid{doi={10.1111/j.1467-9868.2005.00532.x}, issn={1369-7412}, mr={2212574}}
\bptok{imsref}%
\end{barticle}
%
\endbibitem\vadjust{\goodbreak}

\bibitem[\protect\citeauthoryear{Yuan and Lin}{2007}]{yuan2007model}
%
\begin{barticle}[mr]
\bauthor{\bsnm{Yuan},~\bfnm{Ming}\binits{M.}} \AND
\bauthor{\bsnm{Lin},~\bfnm{Yi}\binits{Y.}}
(\byear{2007}).
\btitle{Model selection and estimation in the {G}aussian graphical model}.
\bjournal{Biometrika}
\bvolume{94}
\bpages{19--35}.
\bid{doi={10.1093/biomet/asm018}, issn={0006-3444}, mr={2367824}}
\bptok{imsref}%
\end{barticle}
%
\endbibitem

\end{thebibliography}
\end{document}